\documentclass[%
 reprint,
 superscriptaddress,
 amsmath,amssymb,
 aps,
]{revtex4-1}

\usepackage{graphicx}
\usepackage{dcolumn}
\usepackage{bm}
\usepackage{xcolor}
\usepackage{url}

\usepackage{subfigure}
\usepackage{multirow}

\begin{document}


\title{Deeply Virtual Compton Scattering at Future Electron-Ion Colliders}

\author{Gang Xie}
\affiliation{Institute of Modern Physics, Chinese Academy of Sciences, Lanzhou 730000, China}
\affiliation{School of Nuclear Science and Technology, University of Chinese Academy of Sciences, Beijing 100049, China}

\author{Wei Kou}
\affiliation{Institute of Modern Physics, Chinese Academy of Sciences, Lanzhou 730000, China}
\affiliation{School of Nuclear Science and Technology, University of Chinese Academy of Sciences, Beijing 100049, China}

\author{Qiang Fu}
\affiliation{Institute of Modern Physics, Chinese Academy of Sciences, Lanzhou 730000, China}
\affiliation{School of Nuclear Science and Technology, University of Chinese Academy of Sciences, Beijing 100049, China}

\author{Zhenyu Ye}
\email{yezhenyu@uic.edu; Corresponding author}
\affiliation{Department of Physics, University of Illinois, Chicago, IL 60607, USA}

\author{Xurong Chen}
\email{xchen@impcas.ac.cn; Corresponding author}
\affiliation{Institute of Modern Physics, Chinese Academy of Sciences, Lanzhou 730000, China}
\affiliation{School of Nuclear Science and Technology, University of Chinese Academy of Sciences, Beijing 100049, China}


\date{\today}

\begin{abstract}
The study of hadronic structure has been carried out for many years. Generalized parton distribution functions (GPDs) give broad information on the internal structure of hadrons. Combining GPDs and high-energy scattering experiments, we expect yielding three-dimensional physical quantities from experiments.
Deeply Virtual Compton Scattering (DVCS) process is a powerful tool to study GPDs. It is one of the important experiments of Electron Ion Collider (EIC) and Electron ion collider at China (EicC) in the future. 
In the initial stage, the proposed EicC will have $3 \sim 5$ GeV polarized electrons on $12 \sim 25$ GeV polarized protons, with luminosity up to $1 \sim 2 \times 10^{33}$cm$^{-2}$s$^{-1}$. EIC will be constructed in coming years, which will cover the variable c.m.
energies from 30 to 50 GeV, with the luminosity about $10^{33} \sim 10^{34}$cm$^{-2}$s$^{-1}$.
In this work we present a detailed simulation of DVCS to study the feasibility of experiments at EicC and EIC. Referring the method used by HERMES Collaboration, and comparing the model calculations with pseudo data of asymmetries attributed to the DVCS, we obtained a model-dependent constraint on the total angular momentum of up and down quarks in the proton.

\end{abstract}

\maketitle

\section{Introduction}
\label{sec:intro}

In high energy nuclear physics, the internal structure and dynamics of the proton is still not fully understood. 
Although decades have passed since the discovery that the proton
internal structure consisted of quarks \cite{Gell-Mann:1964ewy,Zweig:1964jf,Bloom:1969kc,Breidenbach:1969kd} and gluons (partons) \cite{Pluto:1978tuc,Darden:1978dk,Darden:1979gd,Bienlein:1978bg}, we still know a little about how the partons contribute to the global properties of the proton such as its mass and spin. 
The measurement of the fraction of the proton spin carried by quarks by the European Muon Collaboration (EMC) in
1987 indicated that only small percentages of the proton's spin comes from quarks \cite{EuropeanMuon:1987isl}. The data of nucleon's polarized structure function $g_{1}\left(x_{B}\right)$ in EMC has deviated significantly from the Ellis-Jaffe sum rule \cite{Ellis:1973kp}.
These results created the so-called "spin crisis", or more appropriately, the "spin puzzle". 
The discrepancy has since inspired many intensive experimental and theoretical studies of spin dependent nucleon structure \cite{Filippone:2001ux,Bass:2004xa,Leader:2013jra,Ji:2016djn,Kuhn:2008sy,Aidala:2012mv,Deur:2018roz}. 
It was proposed that the missing fraction of the proton spin comes from the polarized gluon contribution. 
Recent measurements of the polarized gluon density showed that gluons indeed contribute, but could not fill the gap in the spin puzzle \cite{Aidala:2012mv}. 
The orbital angular momenta of the quarks and gluons play an important role in the proton spin.
According to the generator of Lorentz transformation we can define the angular momentum operator in QCD \cite{Ji:1996ek},
\begin{equation}
	J^{i}=\frac{1}{2} \epsilon^{i j k} \int d^{3} x M^{0 j k},
\end{equation}
where $M^{0 j k}$ is the angular momentum density, which can be expressed by the energy-momentum tensor $T^{\mu \nu}$ through
\begin{equation}
	M^{\alpha \mu \nu}=T^{\alpha \nu} x^{\mu}-T^{\alpha \mu} x^{\nu}.
\end{equation}
$T^{\mu \nu}$ has the Belinfante-Improved form and is symmetric, gauge-invariant, and conserved.
It can be divided into gauge-invariant quark and gluon contributions,
\begin{equation}
	T^{\mu \nu}=T_{q}^{\mu \nu}+T_{g}^{\mu \nu},
\end{equation}
and $\vec{J}$ has a gauge-invariant form, $\vec{J}_{\mathrm{QCD}}=\vec{J}_{q}+\vec{J}_{g}$, where
\begin{equation}
	J_{q, g}^{i}=\frac{1}{2} \epsilon^{i j k} \int d^{3} x\left(T_{q, g}^{0 k} x^{j}-T_{q, g}^{0 j} x^{k}\right).
\end{equation}
In pure gauge theory, $\vec{J}_{g}$ is a conserved angular momentum charge by itself, generating spin
quantum numbers for glueballs. We can see that $\vec{J}_{q}$ and $\vec{J}_{g}$ are interaction-dependent. 
To study the orbital angular momentum of the partons, one needs to study beyond one-dimentional parton distributions.

One-dimensional parton distribution functions (PDFs) provide significant informations about the
structure of the proton. Although the PDFs have provided us with much knowledge
on the proton, one-dimensional distributions can not give us a complete picture. 
Therefore, theorists developed a new density function about 30 years ago, which called GPDs. 
GPDs provide information including both transverse spacial and longitudinal momentum distributions.
Besides the momentum fraction, GPDs depend on another
independent variable, the negative value of momentum transfer square $t=-\left(p-p^{\prime}\right)^{2}$ between the initial and final states of a proton. Thus, the GPDs give extensive informations about three-dimensional dynamics of nucleon, which includes the composition of spin and pressure distribution \cite{CLAS:2022syx,Muller:1994ses,Radyushkin:1996nd,Ji:1996nm,Burkardt:2000za,Polyakov:2002yz}.
Similar to the one dimensional PDFs, GPDs include non-polarized and polarized functions.

GPDs, also named as the off-forward PDFs, have attracted a lot of attention since spin
decomposition rule was first proposed \cite{Ji:1996ek}. 
It was proposed to factorize the hard exclusive processes. The corresponding factorization structure functions including the structure of nucleon are the GPDs $H^{q}\left(x_{B}, \xi, t\right)$, $E^{q}\left(x_{B}, \xi, t\right)$, $\widetilde{H}^{q}\left(x_{B}, \xi, t\right)$ and $\tilde{E}^{q}\left(x_{B}, \xi, t\right)$. These functions correspond to the Fourier transform of
the non-diagonal operators \cite{Guidal:2013rya,Muller:1994ses,Ji:1996ek,Ji:1996nm}:
\begin{equation}
	\begin{array}{l}
		\left.\frac{P^{+}}{2 \pi} \int d y^{-} e^{j x_{B} P^{+} y^{-}}\left\langle p^{\prime}\left|\bar{\Psi}_{q}(0) \gamma^{+} \Psi_{q}(y)\right| p\right\rangle\right|_{y^{+}=\vec{y}_{\perp}=0} \\
		=H^{q}\left(x_{B}, \xi, t\right) \bar{N}\left(p^{\prime}\right) \gamma^{+} N(p) \\
		+ E^{q}\left(x_{B}, \xi, t\right) \bar{N}\left(p^{\prime}\right) i \sigma^{+v} \frac{\Delta_{v}}{2 M_{N}} N(p), \\
		\left.\frac{P^{+}}{2 \pi} \int d y^{-} e^{x_{B} P^{+} y^{-}}\left\langle p^{\prime}\left|\bar{\Psi}_{q}(0) \gamma^{+} \gamma^{5} \Psi_{q}(y)\right| p\right\rangle\right|_{y^{+}=\vec{y}_{\perp}=0} \\
		=\widetilde{H}^{q}\left(x_{B}, \xi, t\right) \bar{N}\left(p^{\prime}\right) \gamma^{+} \gamma_{5} N(p) \\
		+ \widetilde{E}^{q}\left(x_{B}, \xi, t\right) \bar{N}\left(p^{\prime}\right) \gamma_{5} \frac{\Delta^{+}}{2 M_{N}} N(p),
	\end{array}
\end{equation}
where $y$ is the coordinate of the two correlated quarks, the $P$ is the average nucleon four-momentum in light-front frame: $P=\left(p+p^{\prime}\right) / 2$ and $\Delta=p^{\prime}-p$. The "+" superscript means
the plus component of four-momentum in light-front frame. 
Each GPD function defined above is for a specified flavor of quark: $H^{q}, E^{q}, \widetilde{H}^{q}, \widetilde{E}^{q}(q=u, d, s, \ldots)$. 
$H^q$ and $\widetilde{H}^{q}$ are spin non-flipped GPD functions and $E^{q}$ and $\widetilde{E}^{q}$ are spin flipped ones. 
The ordinary parton distributions and nucleon form factors are both included in the off-forward parton distributions. 
In $t \rightarrow 0$ and $\xi \rightarrow 0$ limit, we get
\begin{equation}
	\begin{array}{l}
		H(x_{B}, 0, 0)=f_{1}(x_{B}),     \\ 
		\widetilde{H}\left(x_{B}, 0, 0\right)=g_{1}(x_{B}),
	\end{array}
\end{equation}
where $f_{1}(x_{B})$ is quark distribution and $g_{1}(x_{B})$ is quark helicity distribution. 
According to Dirac and Pauli form factors $F_1$, $F_2$ and axial-vector and pseudo-scalar form factor $G_A$, $G_P$, the sum rules are obtained,
\begin{equation}
	\begin{array}{l}
		\int d x_{B} H\left(x_{B}, \xi, t\right)=F_{1}\left(t\right), \\
		\int d x_{B} E\left(x_{B}, \xi, t\right)=F_{2}\left(t\right), \\
		\int d x_{B} \tilde{H}\left(x_{B}, \xi, t\right)=G_{A}\left(t\right), \\
		\int d x_{B} \tilde{E}\left(x_{B}, \xi, t\right)=G_{P}\left(t\right).
	\end{array}
\end{equation}
The most interesting Ji's sum rules related to the nucleon spins are described through GPDs \cite{Ji:1996nm},
\begin{equation}
	\int_{-1}^{1} d x_{B} x_{B}\left[H\left(x_{B}, \xi, t\right)+E\left(x_{B}, \xi, t\right)\right]=A(t)+B(t).
\end{equation}
Then the total spin of the proton can be expressed as:
\begin{equation}
	\begin{array}{l}
		J_{q, g}=\frac{1}{2}\left[A_{q, g}(0)+B_{q, g}(0)\right], \\
		J_{q}+J_{g}=\frac{1}{2},
	\end{array}
\end{equation}
where $A_{q, g}(0)$ gives the momentum fractions carried by quarks and gluons in the nucleon ($A_{q}(0) + A_{g}(0) = 1$), and B-form factor is analogous to the Pauli form factor for the vector current. By extrapolating the sum rule to $t = 0$, one gets $J_{q,g}$. 
The GPDs can be measured in deep-exclusive processes such as DVCS and deeply virtual meson production (DVMP) \cite{Ji:1996ek,Ji:1996nm,Goeke:2001tz,Vanderhaeghen:1999xj,Goloskokov:2005sd,Collins:1996fb,Mankiewicz:1997bk}. 
Both of these processes are exclusive hard scattering processes in lepton-nucleon collisions.
Theoretical research on these topics has been conducted for many years, and many theoretical models and predictions were created by researchers \cite{Ji:1996ek,Ji:1996nm,Radyushkin:1996nd,JeffersonLabHallA:2007jdm,Belitsky:2001ns,Boffi:2007yc,Gross:2022hyw,Ma:2022gty,Bhattacharya:2022xxw,Schoenleber:2022myb,Braun:2022bpn,MorgadoChavez:2022vzz}. During the past 20 years,
the collaborations at HERA and Jefferson Lab (JLab) have spent a lot effort to get information of GPDs from
electro-production of a real photon (DVCS processes) \cite{H1:2001nez,ZEUS:2003pwh,H1:2005gdw,H1:2007vrx,H1:2009wnw,HERMES:2006pre,HERMES:2011bou,CLAS:2006krx,CLAS:2007clm,CLAS:2008ahu,JeffersonLabHallA:2006prd,JeffersonLabHallA:2007jdm,CLAS:2015bqi,Benali:2020vma,JeffersonLabHallA:2022pnx,JeffersonLabHallA:2015dwe,Defurne:2017paw,CLAS:2001wjj,CLAS:2015uuo,CLAS:2014qtk,CLAS:2018bgk,COMPASS:2018pup,Joerg:2016hhs}, such as DESY with H1 \cite{H1:2001nez,H1:2007vrx}, ZEUS \cite{ZEUS:2003pwh} and
HERMES \cite{HERMES:2006pre,HERMES:2011bou}, JLab Halls A \cite{JeffersonLabHallA:2006prd,JeffersonLabHallA:2015dwe,Defurne:2017paw,JeffersonLabHallA:2007jdm,Benali:2020vma,JeffersonLabHallA:2022pnx} and Halls B \cite{CLAS:2001wjj,CLAS:2007clm,CLAS:2015uuo,CLAS:2014qtk,CLAS:2015bqi,CLAS:2018bgk}, and COMPASS \cite{COMPASS:2018pup,Joerg:2016hhs}. 
These experiments have important contributions to our exploration of the internal structure of the proton. 
Although there are many data from above experiments, the data don't have high precision and wide range of kinematic region.
Accurate measurement of the DVCS process is a huge challenge, which requires high luminosity to compensate for very small cross section and good detector design to ensure the
exclusive measurement of the final states. 
Both EicC and EIC are important experiments in the future that will have very high luminosity and excellent detectors for particle detection.
In this work, we discuss the relation of
GPDs and DVCS observables \cite{Ji:1996nm}, and carry out a
Monte-Carlo simulation of DVCS + Bethe-Heithler (BH) events and do a projection to get the statistical
errors of asymmetry observables of DVCS experiments for the future EicC and EIC. 

Since the contribution of GPDs to amplitude is not independent, the acquisition of GPDs from the exclusive reactions is indirect. We need to use the appropriate GPDs model. 
After years of development, there are many theoretical models of GPD, and two of those are based on double distributions (DDs) \cite{Muller:1994ses,Radyushkin:1998es,Radyushkin:1998bz}, one has been given by Vanderhaeghen, Guichon and Guidal, which called VGG model \cite{Goeke:2001tz,Vanderhaeghen:1998uc,Vanderhaeghen:1999xj,Guidal:2004nd}, another was presented by Goloskokov and Kroll called GK model \cite{Goloskokov:2005sd,Goloskokov:2007nt,Goloskokov:2009ia}. By accessing the available experimental data, the researchers examined different GPD models, and show that the data from different experiments can match well with the VGG model calculation \cite{CLAS:2007clm,Guidal:2013rya,JeffersonLabHallA:2015dwe,CLAS:2014qtk,CLAS:2015uuo}. Based on these results, we perform theoretical calculations with VGG model.
In VGG model, the observable $A_{U T}^{\sin (\phi-\phi_s) \cos \phi}$ is more sensitive to the quark total angular momentum in the nucleon than other parameters \cite{Ye:2006gza,Ellinghaus:2005uc,JeffersonLabHallA:2007jdm}. Thus we make a constraint on $J_u$ and $J_d$ by the pseudo data of Transverse Target-Spin Asymmetry (TTSA) $A_{U T}^{\sin (\phi-\phi_s) \cos \phi}$.

The organization of the paper is as follows. The relationship between GPDs and DVCS is illustrated in Sec. {\ref{sec:DVCSGPD}}. The phenomenological parametrization of GPDs is described in Sec. {\ref{sec:VGGmodel}}. The invariant kinematic and final state kinematic distributions of the simulation are shown in Sec. {\ref{sec:KineDistributions}}. The projections of DVCS experiment are shown in Sec. {\ref{sec:DVCSerror}}.  Finally, some discussions and a concise summary is given in Sec. {\ref{sec:summary}}.

\section{Generalized partons distribution and Deeply Virtual Compton Scattering}
\label{sec:DVCSGPD}

\begin{figure}[htbp]
	\centering
	\includegraphics[scale=0.68]{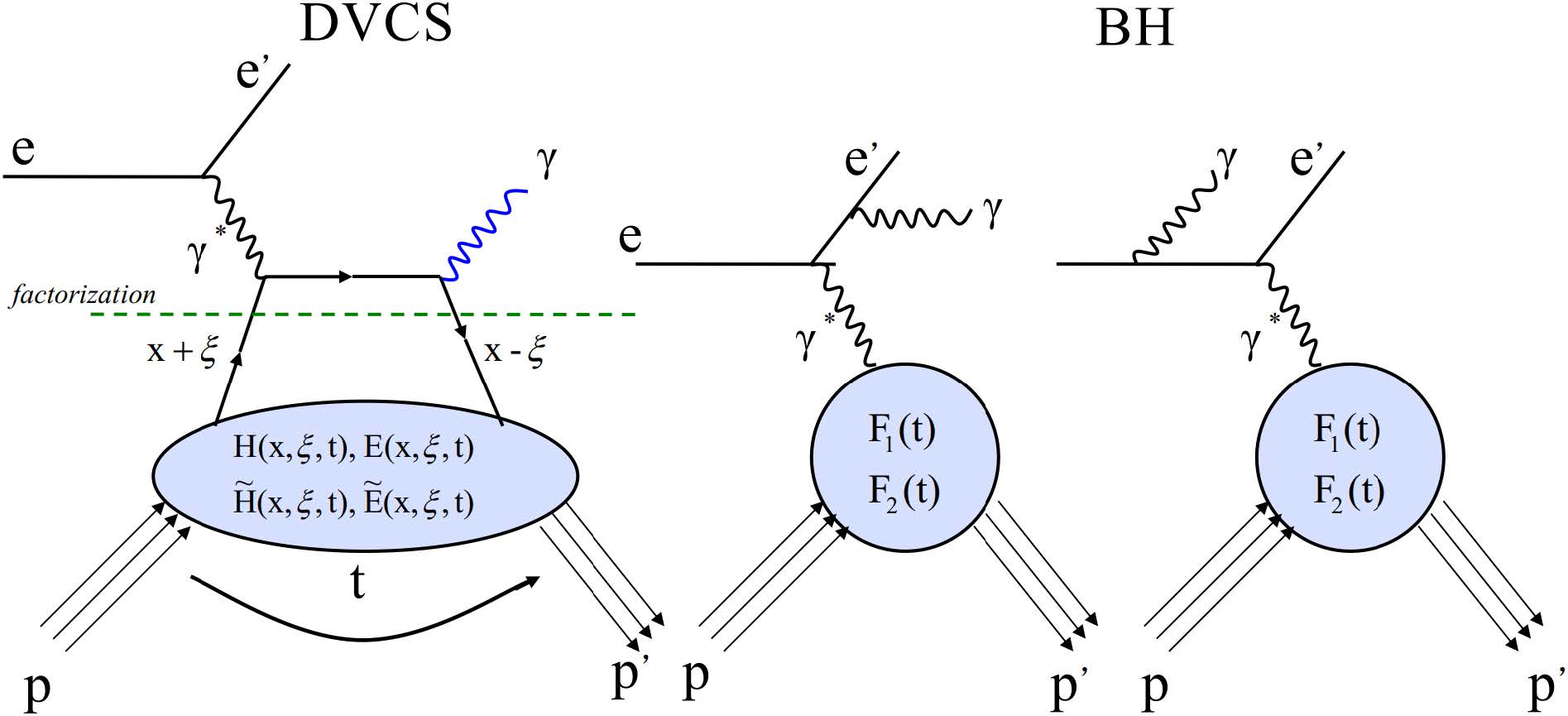}
	\caption{The Feynman diagram of DVCS (left) and BH (right) processes. e, e$^{\prime}$ and p, p$^{\prime}$ are the initial and final states electron and proton respectively. And $t$ is the four-momentum square transition between the initial and final state proton. }
	\label{fig:DVCS-BH-diagram}
\end{figure}

Deeply virtual Compton scattering on a necleon shown in Fig. {\ref{fig:DVCS-BH-diagram}} left panel is the simplest process to access GPDs, it's an important role in exploring the internal structure of necleon. 
In addition to the DVCS, there also
exists another process shares the same final state with DVCS process, see Fig. {\ref{fig:DVCS-BH-diagram}} middle and right panels, called the BH process.

\begin{figure}[htbp]
	\centering
	\includegraphics[scale=0.36]{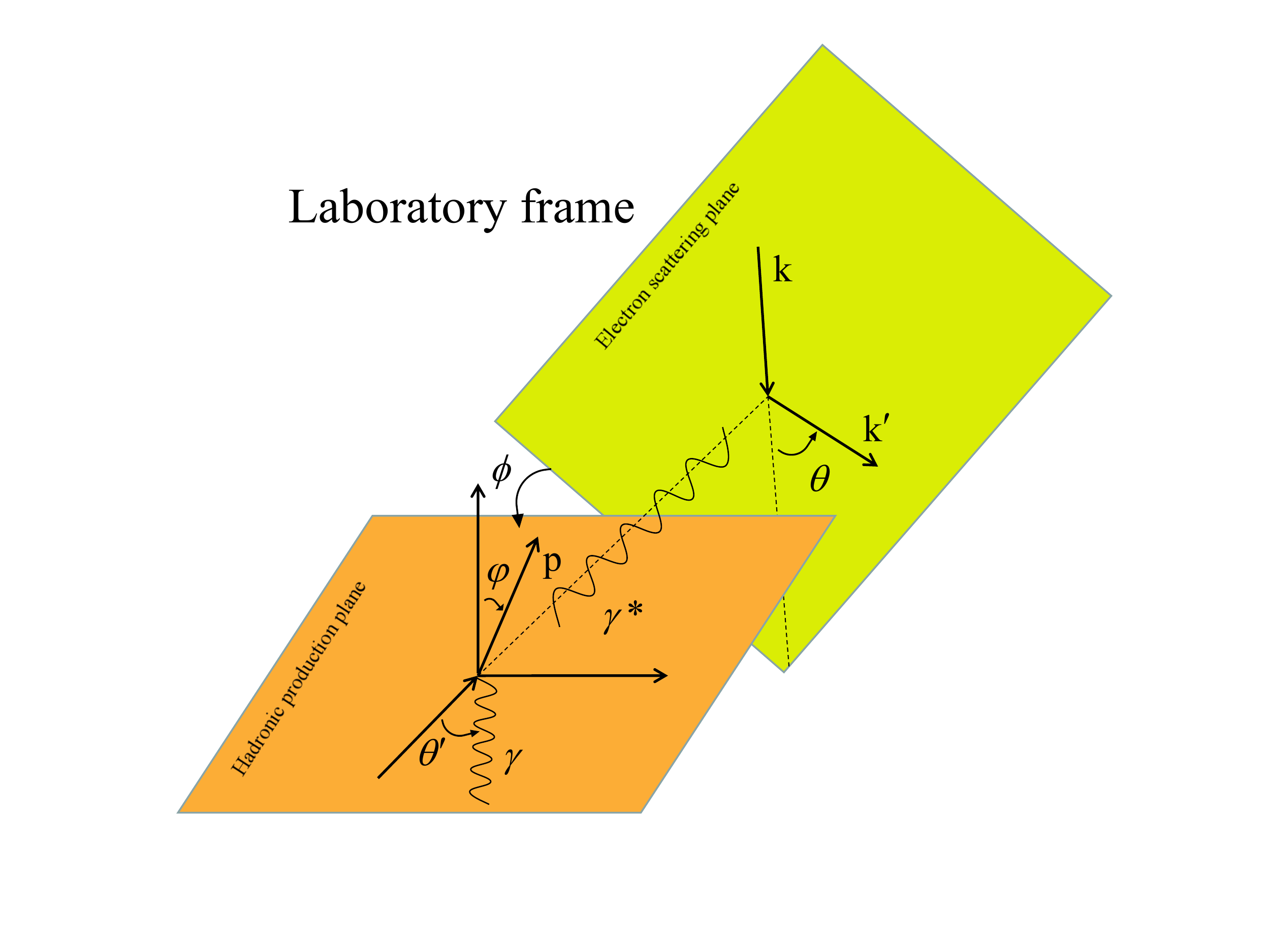}
	\caption{The reference frame of scattering plane and kinematic variables of $e p \rightarrow e^{\prime} p^{\prime} \gamma$ reaction in the
		laboratory \cite{Guidal:2013rya}. }
	\label{fig:DVCS-BH-Process}
\end{figure}

The five-fold differential cross section for electro-production of real photon $e p \rightarrow e^{\prime} p^{\prime} \gamma$ is defined as \cite{Belitsky:2001ns}:
\begin{equation}
	\frac{d \sigma}{d x_{\mathrm{B}} d y d\left|\Delta^{2}\right| d \phi d \varphi}=\frac{\alpha^{3} x_{\mathrm{B}} y}{16 \pi^{2} \mathcal{Q}^{2} \sqrt{1+\epsilon^{2}}}\left|\frac{\mathcal{T}}{e^{3}}\right|^{2}.
\end{equation}
This cross section depends on the common Bjorken scaling variable $x_B$, the squared momentum transfer $ \Delta = \left(P_{2}-P_{1}\right)^{2}$, the lepton energy fraction $y=P_{1} \cdot q_{1} / P_{1} \cdot k$, with $q_{1}=k-k^{\prime}$. The azimuthal angle between the lepton plane and the recoiled proton momentum is $\phi$. There, $\varphi$ is the angle between the polarization vector and the scattered hadron shown in Fig. {\ref{fig:DVCS-BH-Process}}, and $\epsilon=2 x_{\mathrm{B}} M / \mathcal{Q}$ that incorporates nonvanishing target mass effects \cite{Belitsky:2001ns,Belitsky:2012ch}.
The reaction amplitude $\mathcal{T}$ is the linear superposition sum of the BH and DVCS amplitudes,
\begin{equation}
	\mathcal{T}^{2}=\left|\mathcal{T}_{BH}\right|^{2}+\left|\mathcal{T}_{D V C S}\right|^{2}+\mathcal{T}_{I},
\end{equation}
where $\mathcal{T}_{I}=\mathcal{T}_{D V C S} T_{B H}^{*}+\mathcal{T}_{D V C S}^{*} \mathcal{T}_{B H}$. The squared BH term $\left|\mathcal{T}_{BH}\right|^{2}$, squared DVCS amplitude $\left|\mathcal{T}_{D V C S}\right|^{2}$, and interference term $\mathcal{T}_{I}$ are given by:
\begin{equation}
	\begin{aligned}
		\left|\mathcal{T}_{\mathrm{BH}}\right|^{2}= & \frac{e^{6}}{x_{\mathrm{B}}^{2} y^{2}\left(1+\epsilon^{2}\right)^{2} \Delta^{2} \mathcal{P}_{1}(\phi) \mathcal{P}_{2}(\phi)}  \\ 
		 & \left\{c_{0}^{\mathrm{BH}} + \sum_{n=1}^{2} c_{n}^{\mathrm{BH}} \cos (n \phi)+s_{1}^{\mathrm{BH}} \sin (\phi)\right\}, \\
	\end{aligned}
\end{equation}
\begin{equation}
	\begin{aligned}
		& \left|\mathcal{T}_{\mathrm{DVCS}}\right|^{2}= \frac{e^{6}}{y^{2} \mathcal{Q}^{2}} \\
		& \left\{c_{0}^{\mathrm{DVCS}}+\sum_{n=1}^{2}\left[c_{n}^{\mathrm{DVCS}} \cos (n \phi)+s_{n}^{\mathrm{DVCS}} \sin (n \phi)\right]\right\},
	\end{aligned}
\end{equation}
\begin{equation}
	\begin{aligned}
		\mathcal{T}_{I}= & \frac{ \pm e^{6}}{x_{\mathrm{B}} y^{3} \Delta^{2} \mathcal{P}_{1}(\phi) \mathcal{P}_{2}(\phi)} \\
		& \left\{c_{0}^{\mathcal{I}}+\sum_{n=1}^{3}\left[c_{n}^{\mathcal{I}} \cos (n \phi)+s_{n}^{\mathcal{I}} \sin (n \phi)\right]\right\}. 
	\end{aligned}
\end{equation}
The results for the Fourier coefficients can be found in \cite{Belitsky:2001ns,Belitsky:2012ch}. The variables $\xi$ and $t$ (or $\Delta^{2}$) can be computed from the kinematic variables. Since we cannot directly obtain $x_{B}$ from experiment, the Compton form factors (CFF) are obtained by integrating the GPDs,
\begin{equation}
	\begin{aligned}
	& \int_{-1}^{1} \frac{F_{q}(x_{B}, \xi, t)}{x_{B}-\xi+i \epsilon} d x_{B} \\
	& =\mathcal{P} \int_{-1}^{1} \frac{F_{q}(x_{B}, \xi, t)}{x_{B}-\xi} d x_{B}-i \pi F_{q}(\xi, \xi, t),
	\end{aligned}
	\label{eq:CFF}
\end{equation}
where $F_{q}$ are $H^q$, $\widetilde{H}^{q}$, $E^{q}$, or $\widetilde{E}^{q}$. These real and imaginary part of Eq. {\ref{eq:CFF}}, which can be expressed in eight GPD-related quantities that can be extracted from DVCS observables \cite{Guidal:2013rya}:
\begin{equation}
	\begin{aligned}
		H_{R e}(\xi, t) & \equiv \mathcal{P} \int_{0}^{1} d x_{B}\left[H\left(x_{B}, \xi, t\right)-H\left(-x_{B}, \xi, t\right)\right] C^{+}, \\
		H_{I m}(\xi, t) & \equiv H(\xi, \xi, t)-H(-\xi, \xi, t), \\
		E_{R e}(\xi, t) & \equiv \mathcal{P} \int_{0}^{1} d x_{B}\left[E\left(x_{B}, \xi, t\right)-E\left(-x_{B}, \xi, t\right)\right] C^{+}, \\
		E_{I m}(\xi, t) & \equiv E(\xi, \xi, t)-E(-\xi, \xi, t), \\
		\widetilde{H}_{R e}(\xi, t) & \equiv \mathcal{P} \int_{0}^{1} d x_{B}\left[\widetilde{H}\left(x_{B}, \xi, t\right)-\widetilde{H}\left(-x_{B}, \xi, t\right)\right] C^{-}, \\
		\widetilde{H}_{I m}(\xi, t) & \equiv \widetilde{H}(\xi, \xi, t)-\widetilde{H}(-\xi, \xi, t), \\
		\widetilde{E}_{R e}(\xi, t) & \equiv \mathcal{P} \int_{0}^{1} d x_{B}\left[\widetilde{E}\left(x_{B}, \xi, t\right)-\widetilde{E}\left(-x_{B}, \xi, t\right)\right] C^{-}, \\
		\widetilde{E}_{I m}(\xi, t) & \equiv \widetilde{E}(\xi, \xi, t)-\widetilde{E}(-\xi, \xi, t).
	\end{aligned}
\end{equation}
The case with subscript "$Re$" is accessed by observables sensitive to the
real part of the DVCS amplitude, while the case with subscript "$Im$" is accessed by observables sensitive to its imaginary part, where the coefficient $C^{\pm}$ defined as:
\begin{equation}
	C^{\pm}=\frac{1}{x_{B}-\xi} \pm \frac{1}{x_{B}+\xi}.
	\label{eq:coefficientC}
\end{equation}
As a result, the Compton form factors with four complex functions are written as:
\begin{equation}
	\begin{aligned}
		\mathcal{H}(\xi, t) & \equiv H_{R e}(\xi, t)-i \pi H_{I m}(\xi, t), \\
		\tilde{\mathcal{H}}(\xi, t) & \equiv \widetilde{H}_{R e}(\xi, t)-i \pi \widetilde{H}_{I m}(\xi, t), \\
		\mathcal{E}(\xi, t) & \equiv E_{R e}(\xi, t)-i \pi E_{I m}(\xi, t), \\
		\tilde{\mathcal{E}}(\xi, t) & \equiv \widetilde{E}_{R e}(\xi, t)-i \pi \widetilde{E}_{I m}(\xi, t) .
	\end{aligned}
\end{equation}
For the measurement of CFFs, it is mandatory to consider the interference term from BH events. The production of BH events is a pure QED process, which can be measued
precisely from the form factor $F_1$ and $F_2$. 
In addition to the absolute cross section, another way to obtain the CFF is by measuring the asymmetries. 
The beam charge asymmetries are defined as:
\begin{equation}
	A_{C}=\frac{\sigma^{+}(\phi)-\sigma^{-}(\phi)}{\sigma^{+}(\phi)+\sigma^{-}(\phi)},
\end{equation}
where $\sigma^{+}$ and $\sigma^{-}$ refer to cross sections with lepton beams of opposite charge. We can see that the asymmetries only depends on $\phi$. The observables of interest in this paper are the
correlated charge and transversely polarized target-spin asymmetries, defined as:
\begin{equation}
	\begin{array}{l}
		A_{U T, D V C S}=\frac{\left(\sigma_{+}^{+}(\phi)-\sigma_{-}^{+}(\phi)\right)+\left(\sigma_{+}^{-}(\phi)-\sigma_{-}^{-}(\phi)\right)}{\sigma_{+}^{+}(\phi)+\sigma_{-}^{+}(\phi)+\sigma_{+}^{-}(\phi)+\sigma_{-}^{-}(\phi)}, \\
		A_{U T, I}=\frac{\left(\sigma_{+}^{+}(\phi)-\sigma_{-}^{+}(\phi)\right)-\left(\sigma_{+}^{-}(\phi)-\sigma_{-}^{-}(\phi)\right)}{\sigma_{+}^{+}(\phi)+\sigma_{-}^{+}(\phi)+\sigma_{+}^{-}(\phi)+\sigma_{-}^{-}(\phi)},
	\end{array}
\end{equation}
where $A$ with subscripts denote the cross section asymmetries of $e p \rightarrow e^{\prime} p^{\prime} \gamma$ at certain beam
(first subscript) and target (second subscript) polarization sign ("U" stands for unpolarized and "T" for transverse polarized). 
Note that there are two independent transverse polarization direction of proton: $UT_x$ is in the hadronic plane and $UT_y$ is perpendicular to it.
There, the uperscript and subscript of $\sigma$ refers to the charge of the lepton beam and beam (or target) spin projection.  One can
measure exclusive $e p \rightarrow e^{\prime} p^{\prime} \gamma$ cross section with different beam and target polarization since the spin asymmetries give the access to different CFFs through the interference term $\mathcal{I}$, the BH and
DVCS process. 
At leading-order and leading-twist, the relation linking observables and CFFs for $e p \rightarrow e^{\prime} p^{\prime} \gamma$ process have been derived as \cite{Belitsky:2001ns,Roche:2006ex,Kroll:2012sm}:
\begin{equation}
	A_{\mathrm{UT}, \text { DVCS }}^{\sin \left(\phi-\phi_{s}\right)} \propto\left[\operatorname{Im}\left(\mathcal{H} \mathcal{E}^{*}\right)-\xi \operatorname{Im}\left(\widetilde{\mathcal{H}} \widetilde{\mathcal{E}}^{*}\right)\right],
\end{equation}
\begin{equation}
	\begin{aligned}
		A_{\mathrm{UT}, \mathrm{I}}^{\sin \left(\phi-\phi_{s}\right) \cos \phi} \propto & \operatorname{Im}\left[-\frac{t}{4 M^{2}}\left(F_{2} \mathcal{H}-F_{1} \mathcal{E}\right)\right. \\
		& +\xi^{2}\left(F_{1}+\frac{t}{4 M^{2}} F_{2}\right)(\mathcal{H}+\mathcal{E}) \\
		& \left.-\xi^{2}\left(F_{1}+F_{2}\right)\left(\tilde{\mathcal{H}}+\frac{t}{4 M^{2}} \widetilde{\mathcal{E}}\right)\right] .
	\end{aligned}
\end{equation}
These approximations illustrate that different experimental observables are sensitive to different CFFs. We can see that the above asymmetries have dependence on CFF $\mathcal{E}$, which is important implication for our following study of the total angular momentum of different quarks within the proton.

\section{Phenomenological parametrization of GPDs}
\label{sec:VGGmodel}

Assuming a factorized t-dependence, the quark GPD $H^q$
is given by \cite{Goeke:2001tz}:
\begin{equation}
	H^{q}(x, \xi, t)=H^{q}(x, \xi) \cdot F_{1}^{q}(t).
\end{equation}
The nucleon form factors in dipole form is given by:
\begin{equation}
	F_{1}^{\text {dipole }}(t)=\frac{1-\left(1+\kappa^{P}\right) t / 4 m_{N}^{2}}{1-t / 4 m_{N}^{2}} \frac{1}{(1-t / 0.71)^{2}}.
\end{equation}
For the function $H^q$
(for each flavor $q$), the t-independent part $H^{q}(x, \xi) \equiv H^{q}(x, \xi, t=0)$
is parametrized by a two-component form,
\begin{equation}
	H^{q}(x, \xi) \equiv H^{q}_{DD}(x, \xi, t=0) + \theta(\xi-|x|) D^{q}\left(\frac{x}{\xi}\right),
	\label{eq:profun}
\end{equation}
where $D^{q}\left(\frac{x}{\xi}\right)$ is the D-term, set to 0 in our following calculation. And $H^q_{DD}$ is the part of the GPD which is obtained as a one-dimensional section of a
two-variable double distribution (DD) $F^q$, imposing a particular dependence on the skewedness $\xi$,
\begin{equation}
	H_{D D}^{q}(x, \xi)=\int_{-1}^{1} d \beta \int_{-1+|\beta|}^{1-|\beta|} d \alpha \delta(x-\beta-\alpha \xi) F^{q}(\beta, \alpha).
	\label{eq:HDD}
\end{equation}
For the double distributions, entering Eq. \ref{eq:HDD}, we use the following model,
\begin{equation}
	F^{q}(\beta, \alpha)=h(\beta, \alpha) q(\beta),
\end{equation}
where $q(\beta)$ is the forward quark distribution (for the flavor $q$) and where $h(\beta, \alpha)$ denotes
a profile function. In the following estimates, we parametrize the profile function through
a one-parameter ansatz, following \cite{Goeke:2001tz,Radyushkin:1998es,Radyushkin:1998bz}:

\begin{equation}
	h(\beta, \alpha)=\frac{\Gamma(2 b+2)}{2^{2 b+1} \Gamma^{2}(b+1)} \frac{\left[(1-|\beta|)^{2}-\alpha^{2}\right]^{b}}{(1-|\beta|)^{2 b+1}}.
	\label{eq:profilefunction}
\end{equation}
For $\beta > 0$, $q(\beta)=q_{\mathrm{val}}(\beta)+\bar{q}(\beta)$ is the ordinary PDF for the quark flavor $q$. In this work, we use IMParton as input \cite{Wang:2016sfq}. The
negative $\beta$ range corresponds to the antiquark density: $q(-\beta)=-\bar{q}(\beta)$. The parameter $b$ characterizes to what extent the GPD depends on the skewness $\xi$, and fixed to 1 in this work.

The spin-flip quark GPDs $E_q$ in the factorized ansatz are given by:
\begin{equation}
	E_{q}(x, \xi, t)=E_{q}(x, \xi) \cdot F_{2}^{q}(t) / \kappa^{q}.
\end{equation}
Here $F_2^q(t)$ denotes the Pauli FF for quark flavor $q$, and is parameterized by:
\begin{equation}
	F_{2}^{q}=\frac{\kappa^{q}}{\left(1-t / 4 m_{p}^{2}\right) \cdot\left(1-t / m_{D}^{2}\right)^{2}},
\end{equation}
where $\kappa_{q}$ is the anomalous magnetic moment of quarks of flavor $q$, $\kappa^{u}=2\kappa^{p}+\kappa^{n}=1.67$, $\kappa^{d}=\kappa^{p}+2\kappa^{n}=-2.03$. Same as
Eq. \ref{eq:profun}, the t-independent part of the quark GPDs, $E_q(x, \xi)$ is defined as:
\begin{equation}
	E_{q}(x, \xi)=E_{q}^{D D}(x, \xi)-\theta(\xi-|x|) D_{q}\left(\frac{x}{\xi}\right).
\end{equation}
The part of the GPD $E$ that can be obtained from the double distribution has a form
analogous to the spin-nonflip case:
\begin{equation}
	E_{q}^{D D}(x, \xi)=\int_{-1}^{1} d \beta \int_{-1+|\beta|}^{1-|\beta|} d \alpha \delta(x-\beta-\alpha \xi) K_{q}(\beta, \alpha),
\end{equation}
there, $K_{q}(\beta, \alpha)$ is given by:
\begin{equation}
	K_{q}(\beta, \alpha)=h(\beta, \alpha) e_{q}(\beta),
\end{equation}
and $e_q(\beta)$ denotes the spin-flip can be written as:
\begin{equation}
	e_{q}(x)=A_{q} \cdot q_{\mathrm{val}}(x)+B_{q} \cdot \delta(x),
\end{equation}
with:
\begin{equation}
	\begin{aligned}
		A_{q} & =\frac{2 J_{q}-M_{q}^{(2)}}{M_{q_{\text {val }}}^{(2)}}, \\
		B_{u} & =2\left[\frac{1}{2} \kappa_{u}-\frac{2 J_{u}-M_{u}^{(2)}}{M_{u_{\text {val }}}^{(2)}}\right], \\
		B_{d} & =\kappa_{d}-\frac{2 J_{d}-M_{d}^{(2)}}{M_{d_{\text {val }}}^{(2)}}.
	\end{aligned}
\end{equation}
By defining the total fraction of the proton momentum carried by the quarks and antiquarks of flavor $q$ as:
\begin{equation}
	M_{2}^{q}=\int_{0}^{1} d x x[q(x)+\bar{q}(x)]=\int_{0}^{1} d x x\left[q_{\text {val }}(x)+2 \bar{q}(x)\right],
\end{equation}
and the momentum fraction carried by the valence quarks as:
\begin{equation}
	M_{2}^{q_{v a l}}=\int_{0}^{1} d x x q_{\text {val }}(x).
\end{equation}
The parameterizations of $\widetilde{H}$ and $\widetilde{E}$ are introduced in \cite{Goeke:2001tz,Vanderhaeghen:1998uc,Vanderhaeghen:1999xj,Guidal:2004nd}. While parameterization of $\widetilde{H}$, we use polIMParton as input \cite{Han:2021dkc}. In this model, the total angular momentum carried by u-quarks and d-quarks, $J_u$
and $J_d$, are free parameters in the parameterization of the spin-flip GPD $E_q(x,\xi,t)$. Therefore, this parameterization can be used to study the sensitivity of hard electroproduction observables to variations in $J_u$ and $J_d$.

\section{Distributions of invariant and final-state kinematics}
\label{sec:KineDistributions}
There is a package of Monte-Carlo (MC) simulations of DVCS and BH processes called MILOU \cite{Perez:2004ig}. We use this software to generate 5 million events of EicC and EIC. We use the PARTONS (PARtonic Tomography Of Nucleon
Software) package as the observables input \cite{Berthou:2015oaw}. Thus, we can make some pseudo data for subsequent theoretical calculations.
We focus on two future experiments (EIC and EicC), and assume the beam energy of incoming electron and incoming proton with $E_{e}=3.5$ GeV, $E_{p}=20$ GeV at EicC \cite{Anderle:2021wcy}, $E_{e}=5$ GeV, $E_{p}=100$ GeV at EIC \cite{AbdulKhalek:2021gbh}.
We propose to do the measurement of spin azimuthal asymmetries in deeply virtual Compton scattering on  transverse polarized proton. Besides the scattered electron,
real photon and the scattered proton will be measured after the incoming unpolarized
electron. Transverse Target-Spin Asymmetry ($A_{U T}^{\sin (\phi-\phi_s) \cos \phi}$) will be extracted from the data. The EicC facility can
offer the beam integrated luminosity up to 50 fb$^{-1}$, which corresponds to the effective running time within one year \cite{Anderle:2021wcy}. EicC also has a large kinematic acceptance capacity, which can complement the current vacant data. Compared to EicC, EIC offer the beam integrated luminosity up to 60 fb$^{-1}$ in less running time \cite{AbdulKhalek:2021gbh,Aguilar:2019teb}. Combining with the EIC and EicC experiments, high precision data of most kinematic regions will be availabled.

\begin{figure}[htbp]
	\centering
	\includegraphics[scale=0.36,angle=0]{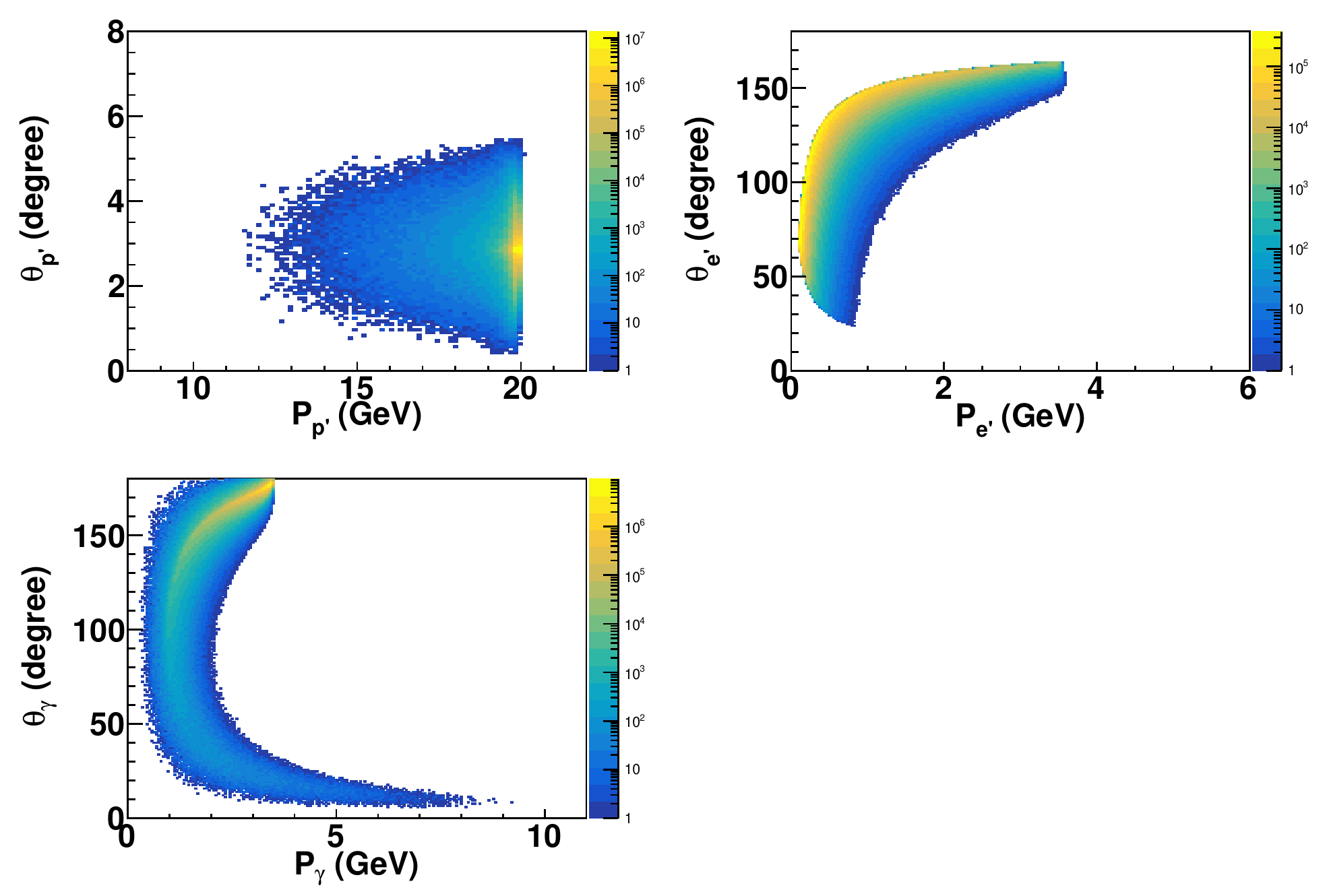}
	\caption{The cross-section weighted momentum and polar angles distributions
		of the final-state particles (scattered protons, scattered electrons and real photons) in the MC simulation at EicC.}
	\label{fig:finalstate_EicC}
\end{figure}

\begin{figure}[htbp]
	\centering
	\includegraphics[scale=0.36,angle=0]{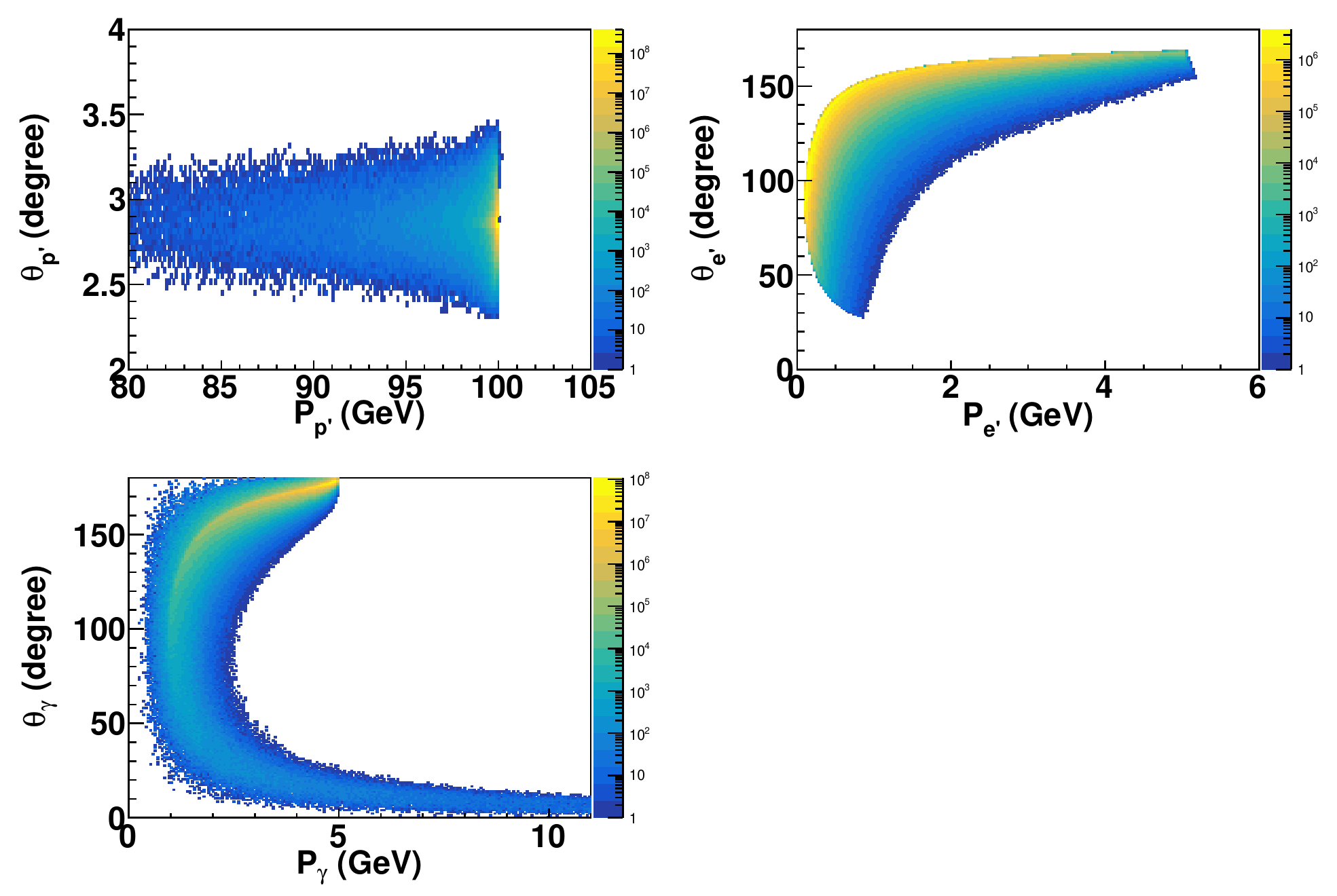}
	\caption{The cross-section weighted momentum and polar angles distributions
		of the final-state particles (scattered protons, scattered electrons and real photons) in the MC simulation at EIC.}
	\label{fig:finalstate_EIC}
\end{figure}

\begin{figure}[htbp]
	\centering
	\includegraphics[scale=0.36,angle=0]{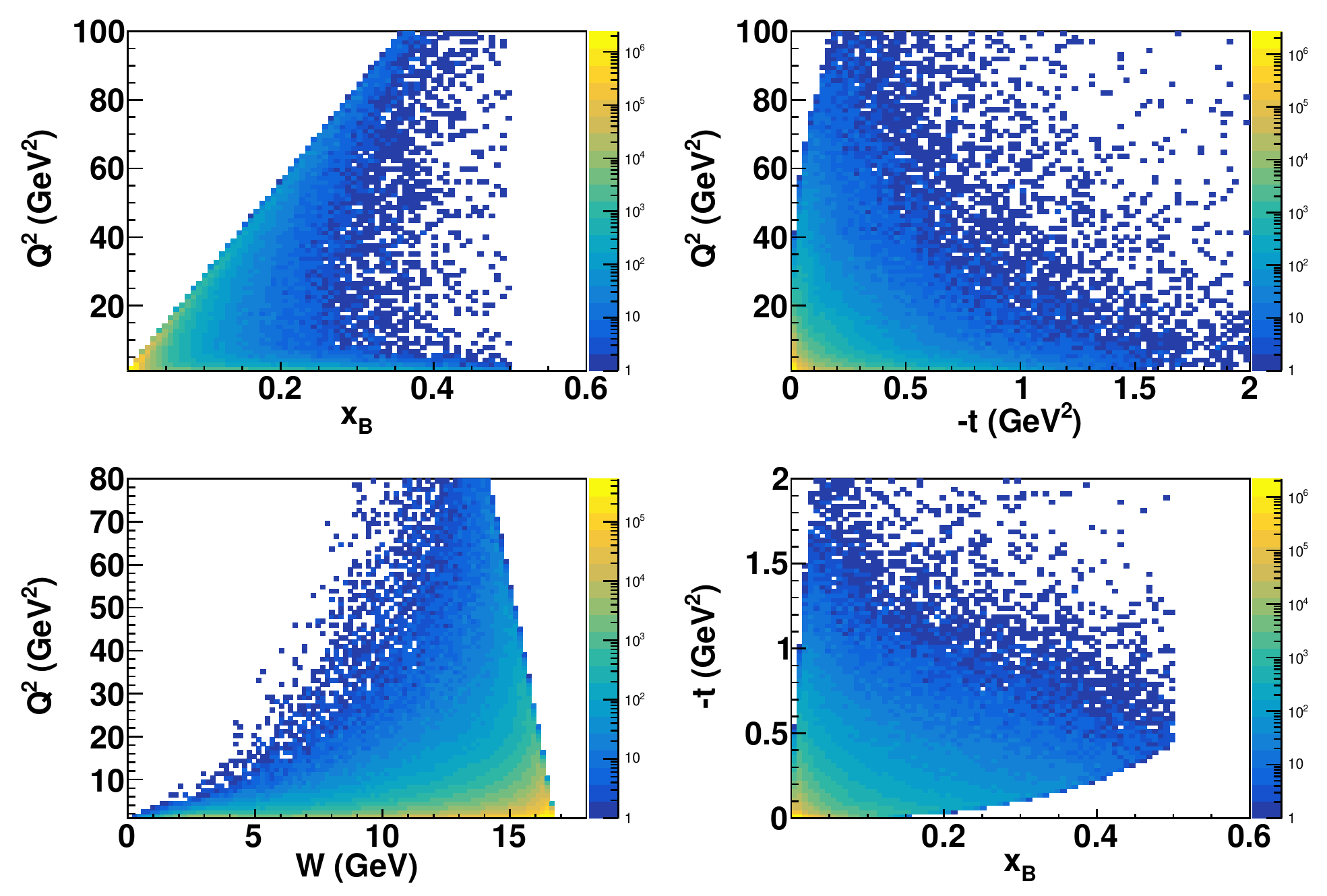}
	\caption{The cross-section weighted distributions of the invariant kinematics in the MC simulation at EicC.}
	\label{fig:kinvar_EicC}
\end{figure}

\begin{figure}[htbp]
	\centering
	\includegraphics[scale=0.36,angle=0]{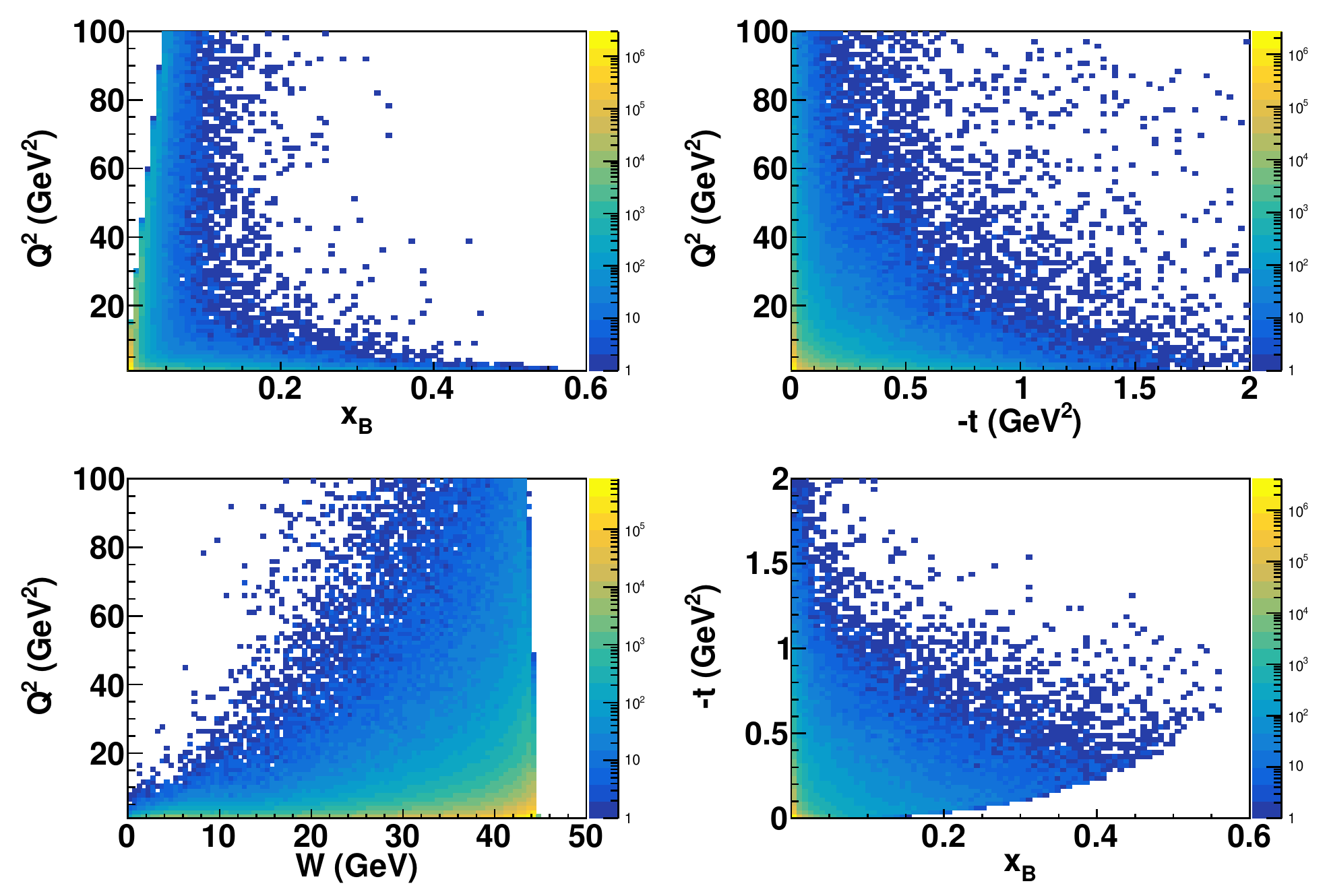}
	\caption{The cross-section weighted distributions of the invariant kinematics in the MC simulation at EIC.}
	\label{fig:kinvar_EIC}
\end{figure}

In order to efficiently generate
the events in the kinematic region of interests, we apply the following kinematical ranges for the Monte-Carlo sampling: $10^{-4}<x_{B}<1$, 1 GeV$^2$$<Q^2<$ 100 GeV$^2$, and $10^{-3}$ GeV$^2$$<-t<$ 3 GeV$^2$.
Fig. \ref{fig:finalstate_EicC} and Fig \ref{fig:finalstate_EIC} show the coverage of the momentum vs polar angles for final state electrons, real photons and scattered protons coming from DVCS and BH process at EicC and EIC. 
We see that the final proton having a large fraction of the momentum of the incoming proton and a small scattering angle. Especially, most protons locate at very small polar angles, and the momentum difference with beam is so small that we need very good momentum resolution for the forward detector. The final electron having a larger scattering angle than the final proton. 
According to the distribution of the final state particles, we can place the detectors appropriately to collect more valid examples. 
Fig. \ref{fig:kinvar_EicC} and Fig. \ref{fig:kinvar_EIC} show the cross-section weighted invariant kinematics distributions of $e p \rightarrow e^{\prime} p^{\prime} \gamma$ reaction at EicC and EIC.
These color z-axis distribution were weighted by the cross section computed in VGG model built in the MILOU software and shown in Log z scale. 
We can see that, the range of $Q^2$ covers from 1.0 GeV$^2$ to 10.0 GeV$^2$, $x_B$
lies between 0.003 and 0.05, and $t$ goes from 0 down to -0.2 GeV$^2$, most of the events are in this area. 
Comparing the results of EicC and EIC, we can see that EIC has more data in the smaller $x_B$ and smaller $-t$ region than EicC.

\section{PROJECTION OF DVCS EXPERIMENT}
\label{sec:DVCSerror}
\begin{figure}[htbp]
	\centering
	\includegraphics[scale=0.36,angle=0]{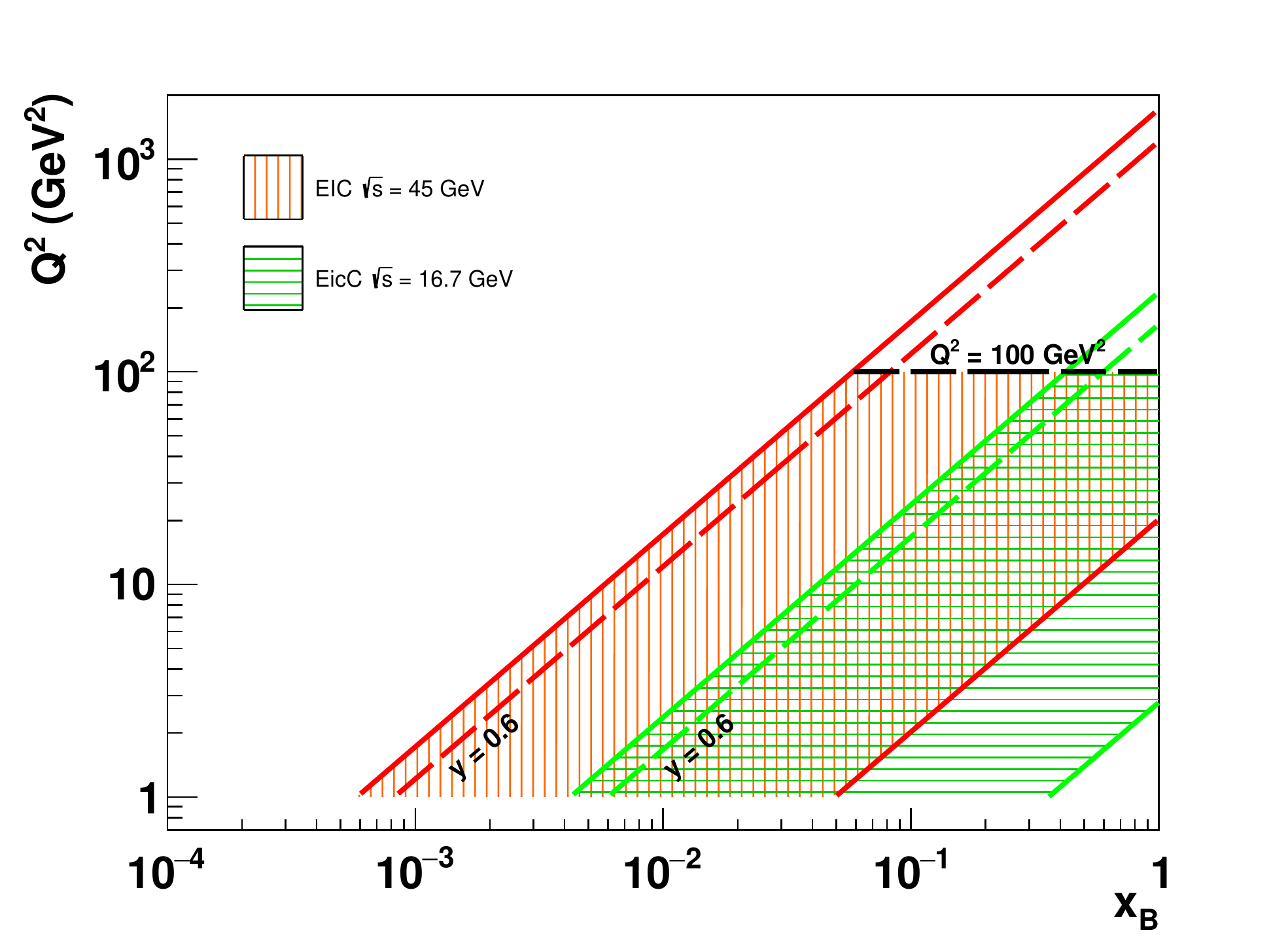}
	\caption{Kinematic range in the $x$, $Q^2$ plane at EicC ($\sqrt{s}=16.7$ GeV) and EIC ($\sqrt{s}=45$ GeV) \cite{Aschenauer:2012ve,Accardi:2012qut,Cao:2023wyz}. The hatched areas indicate the areas simulated in this work, which correspond to $0.01 \leq y \leq 0.85$. The red dashed line and green dashed line indicate $y = 0.6$.}
	\label{fig:EicC_EIC_region}
\end{figure}

\begin{figure}[htbp]
	\centering
	\includegraphics[scale=0.36,angle=0]{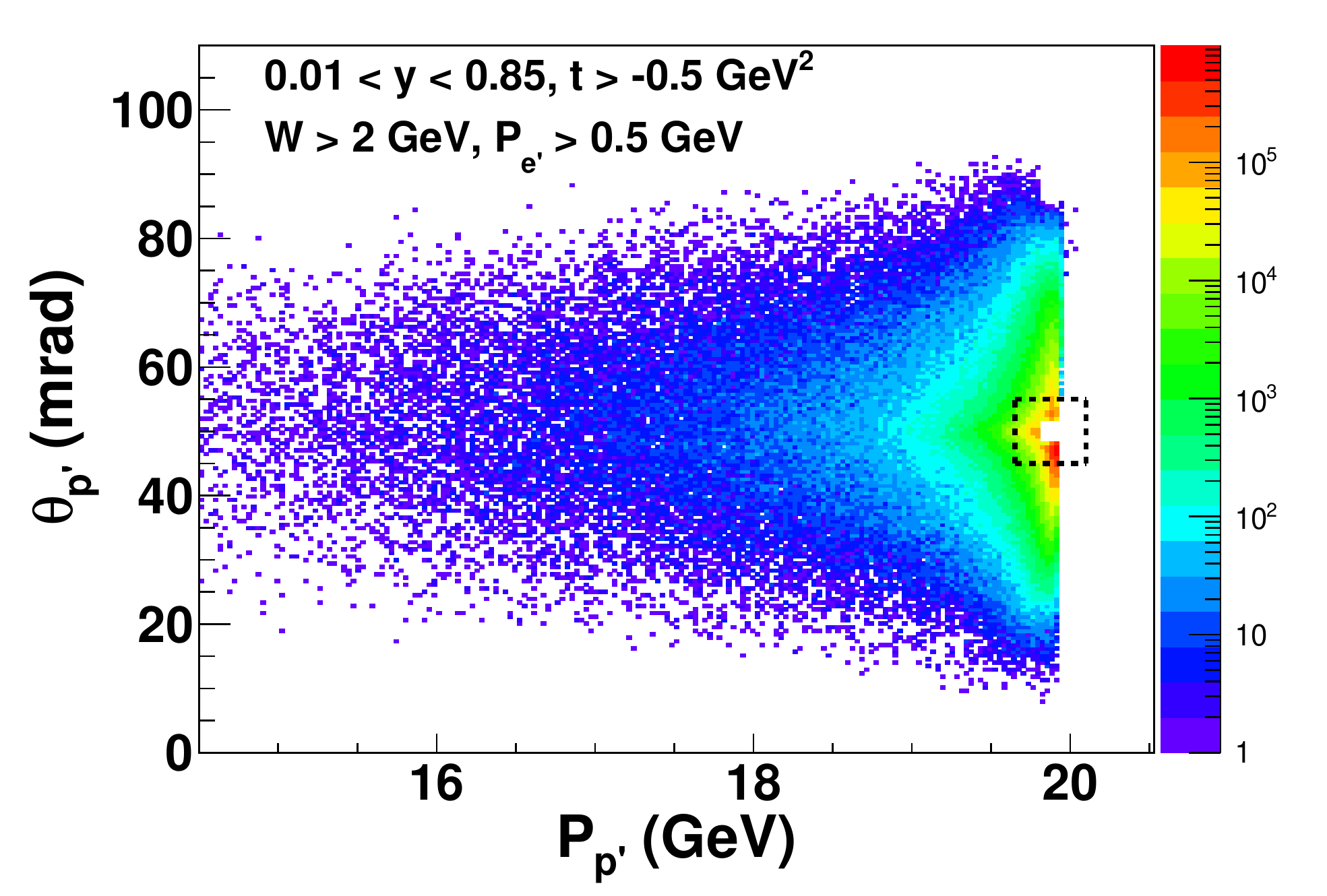}
	\caption{The cross-section weighted momentum and polar angles distributions
		of the scattered protons with the geometric cut. The square breach at the right side shows the eliminated data with proton momentum larger than 99 $\%$ of beam momentum and scattering angle smaller than 2 mrad.}
	\label{fig:protoncut}
\end{figure}

The statistical uncertainty of the measured experimental observable is directly related to the number of events collected during an experiment.
To estimate the number of events of an experiment, we need to know the cross section of the reaction, the integrated luminosity of the experiment, and the events selection criteria of the reaction.
EIC yields an integrated luminosity of 1.5 fb$^{-1}$ per month \cite{AbdulKhalek:2021gbh}.
We assume the integrated luminosity of the experiment of EicC to be 50 fb$^{-1}$, which takes three to four years. The integrated luminosity of EIC is assumed to be 60 fb$^{-1}$ about three years.
To make sure the collected events are valid for our study, we have applied the following conditions for the event selection: $0.01<\mathrm{y}<0.85$, $t>-0.5$ GeV$^2$, $\mathrm{W}>2.0$ GeV, $\mathrm{P}_{\mathrm{e}^{\prime}}>0.5$ GeV. Fig. \ref{fig:EicC_EIC_region} shows the kinematic regions of EIC and EicC, which is the simulated region in this work. EIC and EicC will provide data in small $x$ region. Red area is indicating EIC and green area is indicating EicC. In the small $Q^2$ region, EicC can provide data, where $x$ is close to $x\sim0.005$. Since EIC has higher center-of-mass energy, it can provide data for more smaller $x$-region in the range of $x\sim0.0007$.
DVCS experiment poses strong challenges to us on the detection of recoiled proton with
small $t$. In order to make sure that the recoiled proton can be detected by forward detector, we assumed some constraints on the detection of final state protons. This low-t acceptance eliminates many forward events, taking EicC as an example (Fig. \ref{fig:protoncut}).

Based on the event selection criteria discussed above, the number of events in each bin is calculated with the following formula,
\begin{equation}
	N=\sigma^{a v g} \cdot \text { Lumi } \cdot \text { Time } \cdot \epsilon_{\mathrm{eff}} \cdot \Delta x_{B} \cdot \Delta t \cdot \Delta Q^{2},
	\label{eq:eventnum}
\end{equation}
where $N$ is the total events in each kinematical bins, $\sigma^{a v g}$ is the average of the four cross section with different electron and proton beam polarization directions, "Lumi" is the beam luminosity, "Time" is the beam duration, and $\epsilon_{\mathrm{eff}}$ is the overall efficiency of detector, and the rest denotes the sizes of the kinematical bins. In this work, we conservatively assumed an acceptance of final state particles, which is 25 $\%$ at EIC and 20 $\%$ at EicC \cite{Anderle:2021wcy,AbdulKhalek:2021gbh}. 

\begin{figure}[htbp]
	\centering
	\includegraphics[scale=0.36,angle=0]{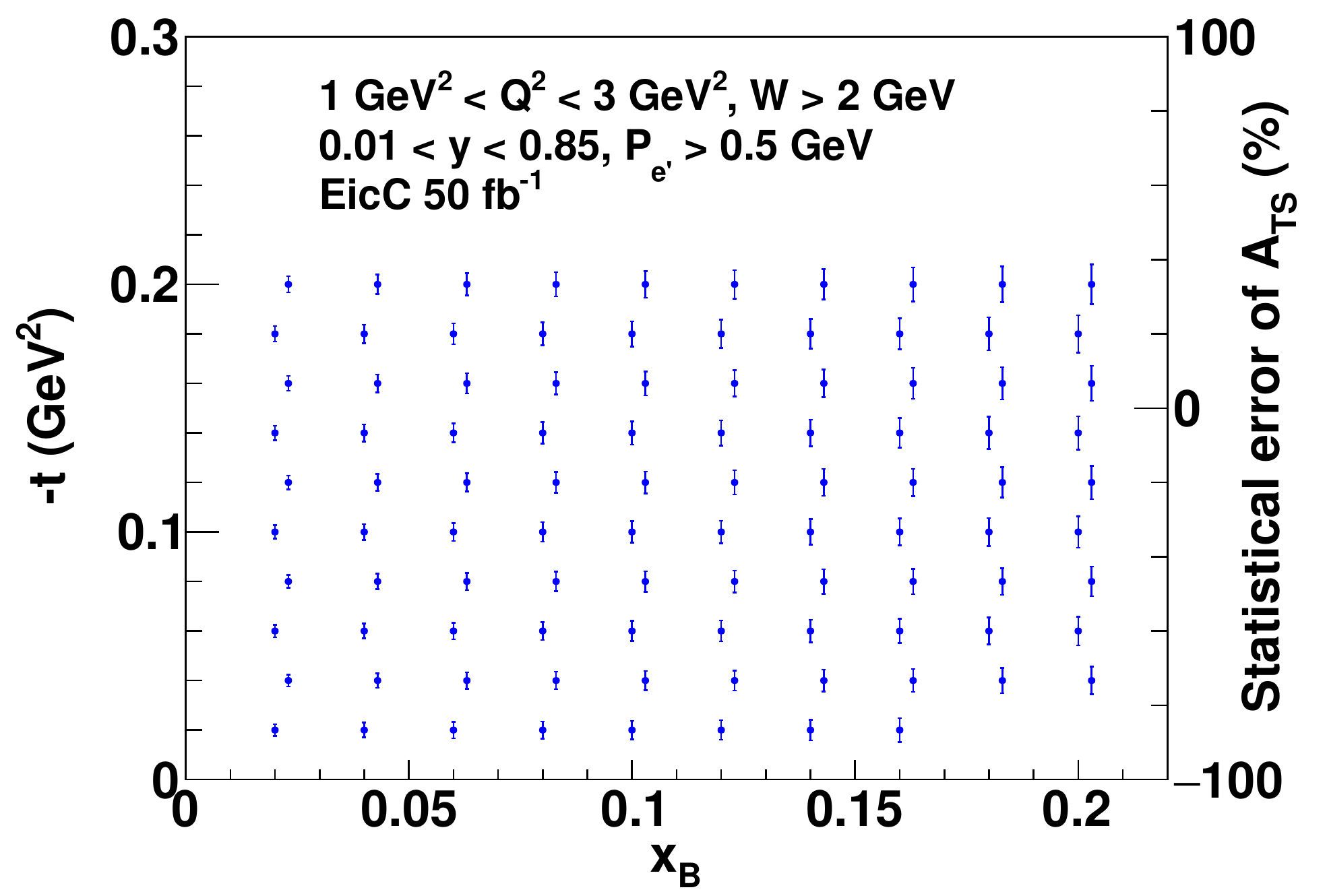}
	\caption{The statistical errors projection of the Transverse Target-Spin Asymmetry at low $Q^2$ at EicC. We calculate the statistical errors at each bin center. The right axis shows how large the statistical errors are.}
	\label{fig:statlowQ2_EicC}
\end{figure}

\begin{figure}[htbp]
	\centering
	\includegraphics[scale=0.36,angle=0]{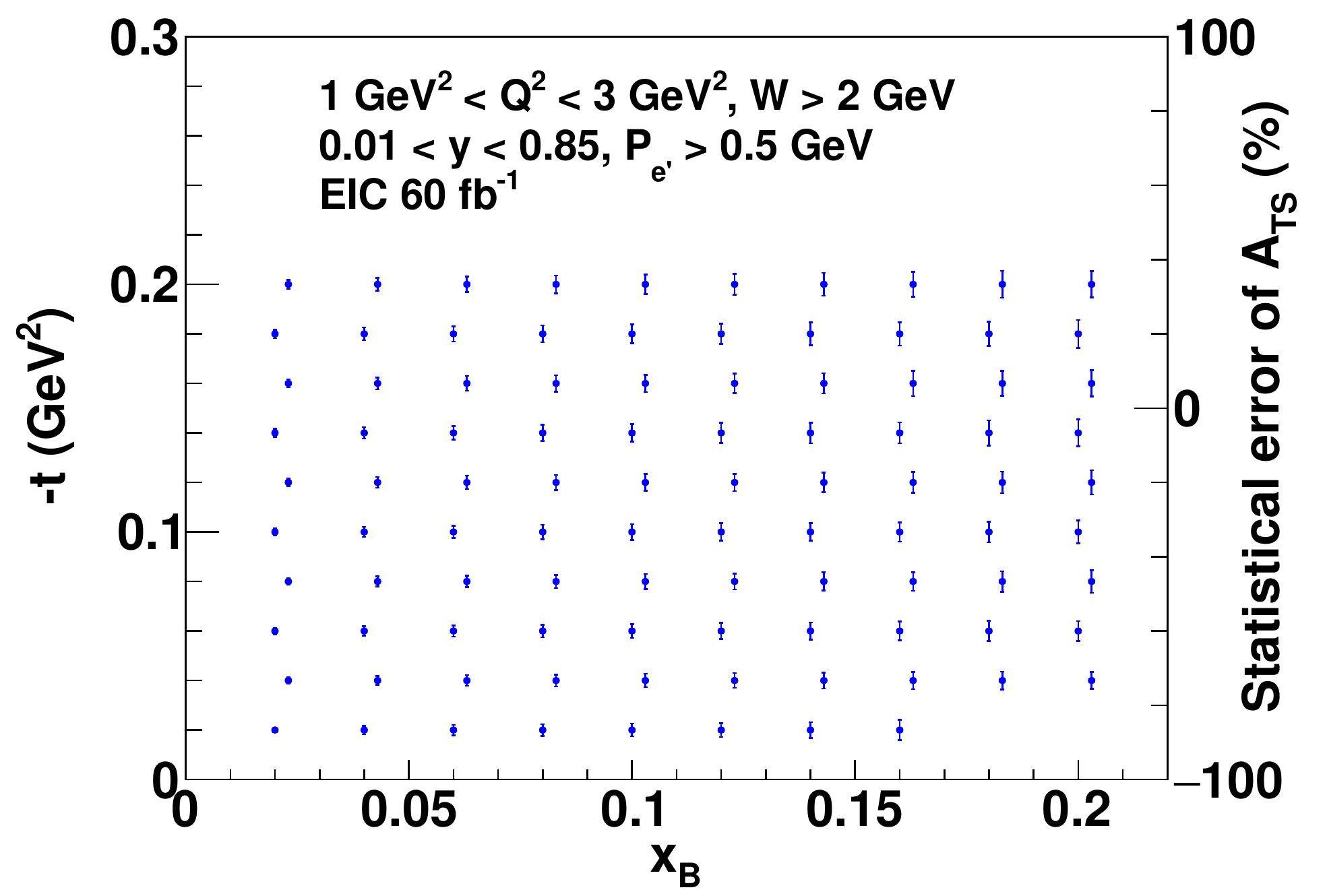}
	\caption{The statistical errors projection of the Transverse Target-Spin Asymmetry at low $Q^2$ at EIC. We calculate the statistical errors at each bin center. The right axis shows how large the statistical errors are.}
	\label{fig:statlowQ2_EIC}
\end{figure}

\begin{figure}[htbp]
	\centering
	\includegraphics[scale=0.36,angle=0]{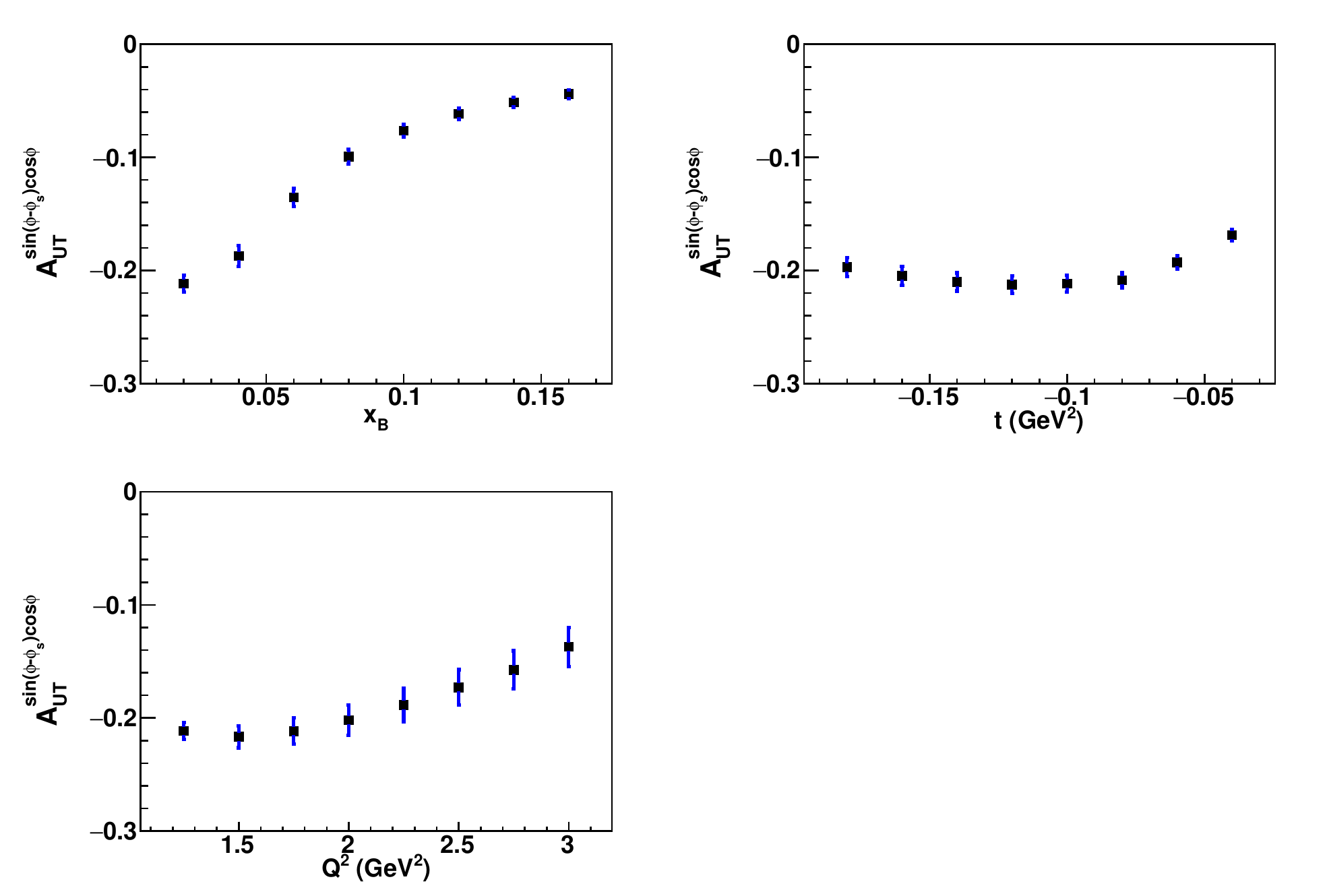}
	\caption{Asymmetries with polarized electron beam and proton beam in some typical bins at EicC.}
	\label{fig:Asystat_EicC}
\end{figure}

\begin{figure}[htbp]
	\centering
	\includegraphics[scale=0.36,angle=0]{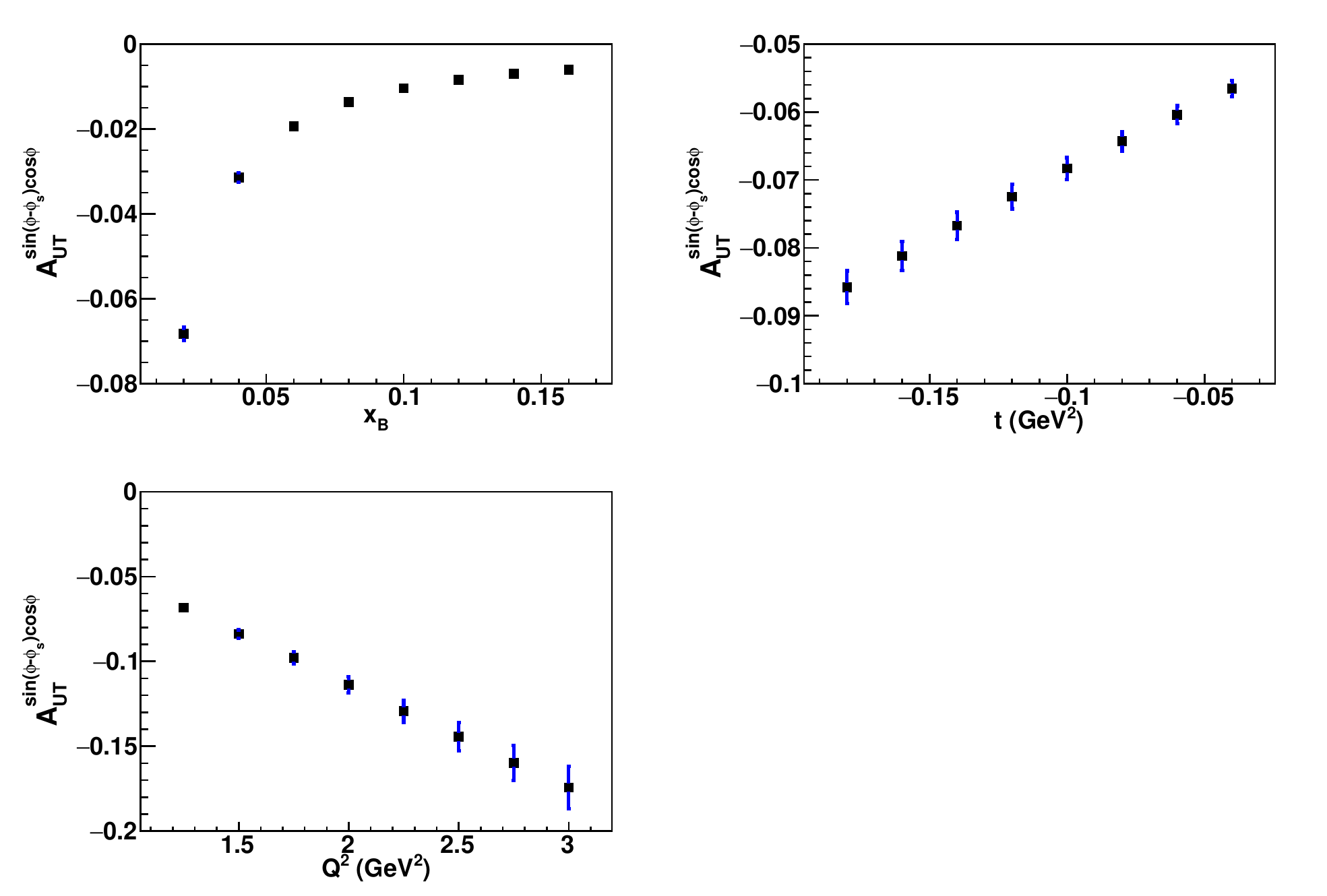}
	\caption{Asymmetries with polarized electron beam and proton beam in some typical bins at EIC.}
	\label{fig:Asystat_EIC}
\end{figure}

The counts of
events in each bin is denoted as $N^{++}$, $N^{+-}$, $N^{-+}$, and $N^{--}$, corresponding to different electron and nucleon polarization directions. One can obtain the asymmetries quantities of the target spin asymmetry ($A_{TS}$):
\begin{equation}
	A_{TS}=\frac{N^{++}+N^{-+}-N^{+-}-N^{--}}{N^{++}+N^{+-}+N^{-+}+N^{--}} \frac{1}{P_{T}},
\end{equation}
where $P_{T}$ stands for the polarization degree of nucleon (assumed
as 70 $\%$) \cite{Anderle:2021wcy,AbdulKhalek:2021gbh}. Considering that the
asymmetries quantities are in several percent level, we use the unpolarized events by MILOU to
do the projection, and the total event number of all polarization conditions is denoted as $N$. Thus
the absolute statistical uncertainty of the asymmetries quantities can be expressed approximately
as:
\begin{equation}
	\delta A_{TS}\approx \frac{1}{P_{T}} \frac{1}{\sqrt{N}}.
\end{equation}
Fig. \ref{fig:statlowQ2_EicC} and Fig. \ref{fig:statlowQ2_EIC} show the statistical errors projection in a low $Q^2$ bin between 1 and 3 GeV$^2$ for EicC and EIC experiments. We focus on small $x_B$ and $-t$ region, and divide the $x_B$ vs. $-t$ plane into very small bins. 
We see in these plots that the statistical uncertainty goes up with $x_B$ increasing. For most of the data at EicC and EIC, the projected statistical uncertainty is smaller than 3 $\%$. When $x_B$ increasing to around 0.12,  the statistical uncertainty is around 5 $\%$. These precise data will be of great help to theoretical research in the future.
Now we can give the pseudo-data of the asymmetry of the cross-section in the area of interest at EicC and EIC. We divide $x_B$, $t$, and $Q^2$ in different bins, show in Tab. \ref{tab:binningscheme}. This table corresponds to the Fig. \ref{fig:Asystat_EicC} and Fig. \ref{fig:Asystat_EIC}. For the case where only $x_B$, $t$ or $Q^2$ changes, we applied a similar division approach. 
Here $x_B$ ranges from 0.01 to 0.17 in steps of 0.02 ($t:-0.11\sim-0.09$ GeV$^2$, $Q^2:1.13\sim1.38$ GeV$^2$), $t$ ranges from -0.19 GeV$^2$ to -0.03 GeV$^2$ 
in steps of 0.02 ($x_B:0.01\sim0.03$, $Q^2:1.13\sim1.38$ GeV$^2$) and
$Q^2$ ranges from 1.13 GeV$^2$ to 3.13 GeV$^2$ in steps of 0.25 ($x_B:0.01\sim0.03$, $t:-0.11\sim-0.09$ GeV$^2$). As shown in Fig. \ref{fig:Asystat_EicC}, EicC provides large phase space coverage and good statistics, especially for small $x_B$,$-t$ and $Q^2$ regions. The similar results at EIC \cite{Burkert:2022hjz} are shown in Fig. \ref{fig:Asystat_EIC}. Since we also divide the $Q^2$ into small bins, the statistical errors of pseudo-data in Fig. \ref{fig:Asystat_EicC} and Fig. \ref{fig:Asystat_EIC} are much larger than those shown in Fig. \ref{fig:statlowQ2_EicC} and Fig. \ref{fig:statlowQ2_EIC}.

\begin{table}
	\caption{
		Binning scheme for $x_B$, $t$, and $Q^2$.
	}
	\label{tab:binningscheme}
	\begin{tabular}{|m{2cm}|m{5em}|m{5em}|m{5em}|}
		\hline
		& $x_B$ & $t$ (GeV$^2$) & $Q^{2}$ (GeV$^2$)\\
		\hline
		$x_B$ & $\quad \quad\backslash$ & 0.01$\sim$0.03 & 0.01$\sim$0.03 \\
		\hline
		$t$ (GeV$^2$) & -0.11$\sim$-0.09 & $\quad \quad\backslash$ & -0.11$\sim$-0.09 \\
		\hline
		$Q^{2}$ (GeV$^2$) & 1.13$\sim$1.38 & 1.13$\sim$1.38 & $\quad \quad\backslash$  \\
		\hline
		bins & 0.01$\sim$0.03 0.03$\sim$0.05 0.05$\sim$0.07 0.07$\sim$0.09 0.09$\sim$0.11 0.11$\sim$0.13 0.13$\sim$0.15 0.15$\sim$0.17 & -0.05$\sim$-0.03 -0.07$\sim$-0.05 -0.09$\sim$-0.07 -0.11$\sim$-0.09 -0.13$\sim$-0.11 -0.15$\sim$-0.13 -0.17$\sim$-0.15 -0.19$\sim$-0.17 & 1.13$\sim$1.38 1.38$\sim$1.63 1.63$\sim$1.88 1.88$\sim$2.13 2.13$\sim$2.38 2.38$\sim$2.63 2.63$\sim$2.88 2.88$\sim$3.13 \\ 
		\hline
	\end{tabular}
\end{table}

\begin{table}
	\caption{
		Asymmetries with polarized electron beam and proton beam at EicC.
	}
	\label{tab:smallxdataEicC}
	\begin{tabular}{|m{4em}|m{4em}|m{5em}|m{10em}|}
	\hline
	$x_B$ & $t$ (GeV$^2$) & $Q^{2}$ (GeV$^2$) & $A_{U T}^{\sin (\phi-\phi_s) \cos \phi}\pm stat$\\
	\hline
	0.006 & 0.10 & 1.25 & -0.089$\pm$0.007 \\
	\hline
	0.01 & 0.10 & 1.25 & -0.168$\pm$0.016 \\ 
	\hline
	0.1 & 0.12 & 2.50 & -0.142$\pm$0.020 \\
	\hline
\end{tabular}
\end{table}

\begin{table}
	\caption{
		Asymmetries with polarized electron beam and proton beam at EIC.
	}
	\label{tab:smallxdataEIC}
	\begin{tabular}{|m{4em}|m{4em}|m{5em}|m{10em}|}
		\hline
		 $x_B$ & $t$ (GeV$^2$) & $Q^{2}$ (GeV$^2$) & $A_{U T}^{\sin (\phi-\phi_s) \cos \phi}\pm stat$\\
		\hline
		0.002 & 0.10 & 1.25 & -0.225$\pm$0.005 \\
		\hline
		0.006 & 0.10 & 1.25 & -0.172$\pm$0.008 \\
		\hline
		0.01 & 0.10 & 1.25 & -0.121$\pm$0.007 \\ 
		\hline
		0.1 & 0.12 & 2.50 & -0.020$\pm$0.002 \\
		\hline
	\end{tabular}
\end{table}

\begin{figure}[htbp]
	\centering
	\includegraphics[scale=0.36,angle=0]{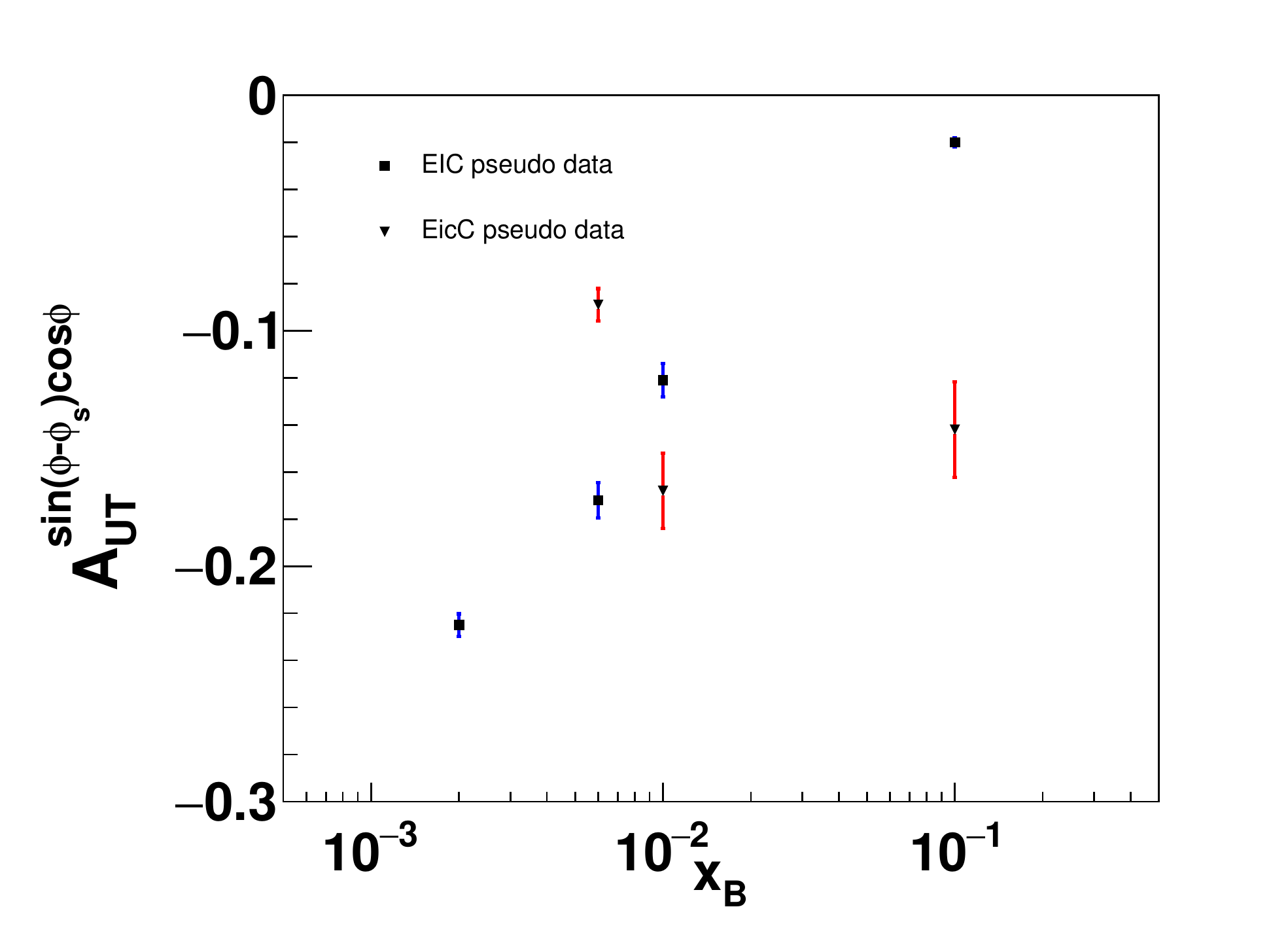}
	\caption{Asymmetries with polarized electron beam and proton beam in small $x$ region at EicC (Tab. \ref{tab:smallxdataEicC}) and EIC (Tab. \ref{tab:smallxdataEIC}).}
	\label{fig:Asystat_EicC_EIC_small_x}
\end{figure}

\begin{figure}[htbp]
	\centering
	\includegraphics[scale=0.46,angle=0]{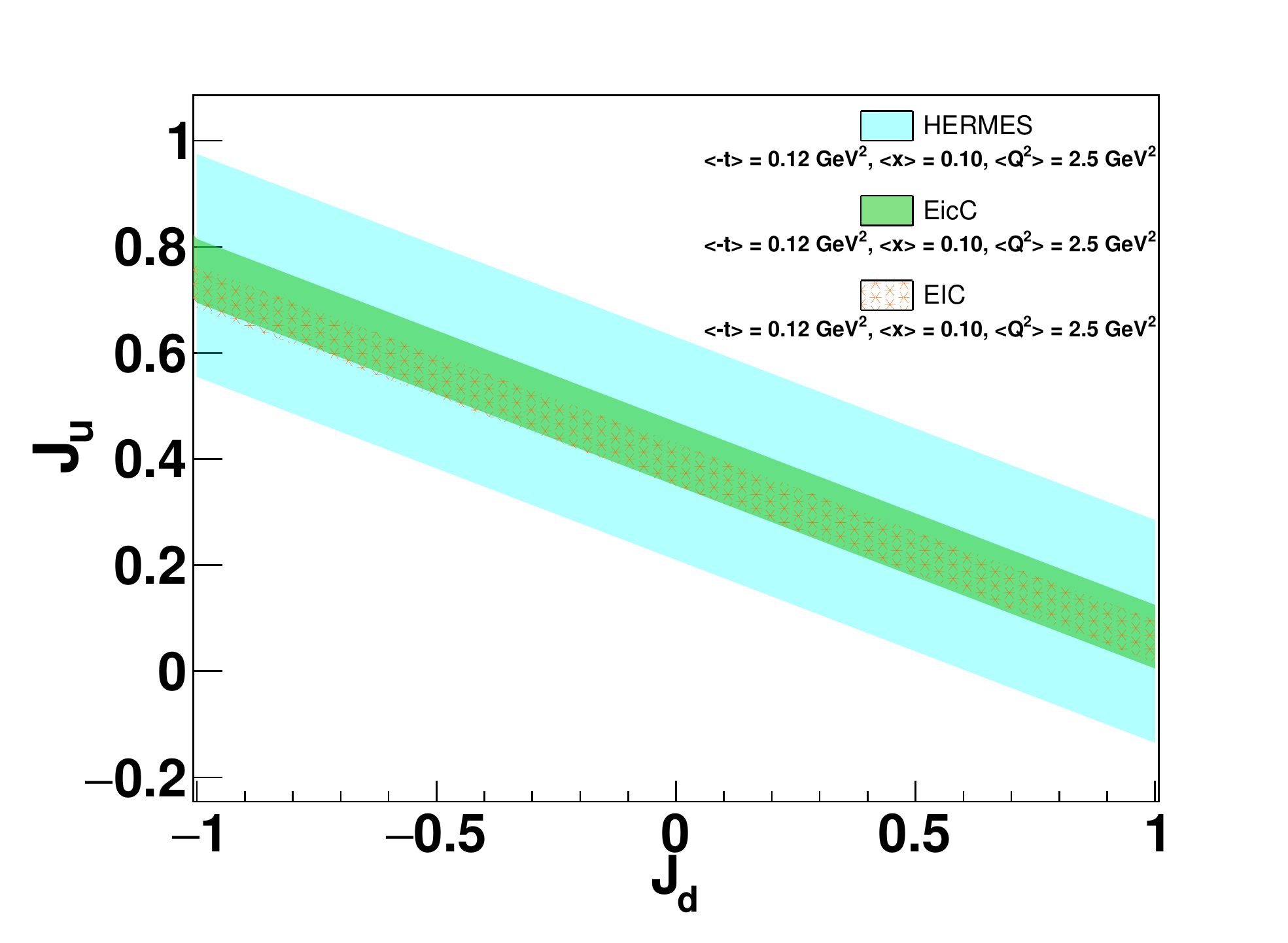}
	\caption{The result of model-dependent constraint on u-quark total angular momentum $J_u$ vs d-quark total angular momentum $J_d$ at EIC and EicC compared with HERMES \cite{Ellinghaus:2005uc,Ye:2006gza}. Only statistical errors are considered. }
	\label{fig:EIC_EicC_HERMESband01}
\end{figure}

\begin{figure}[htbp]
	\centering
	\includegraphics[scale=0.46,angle=0]{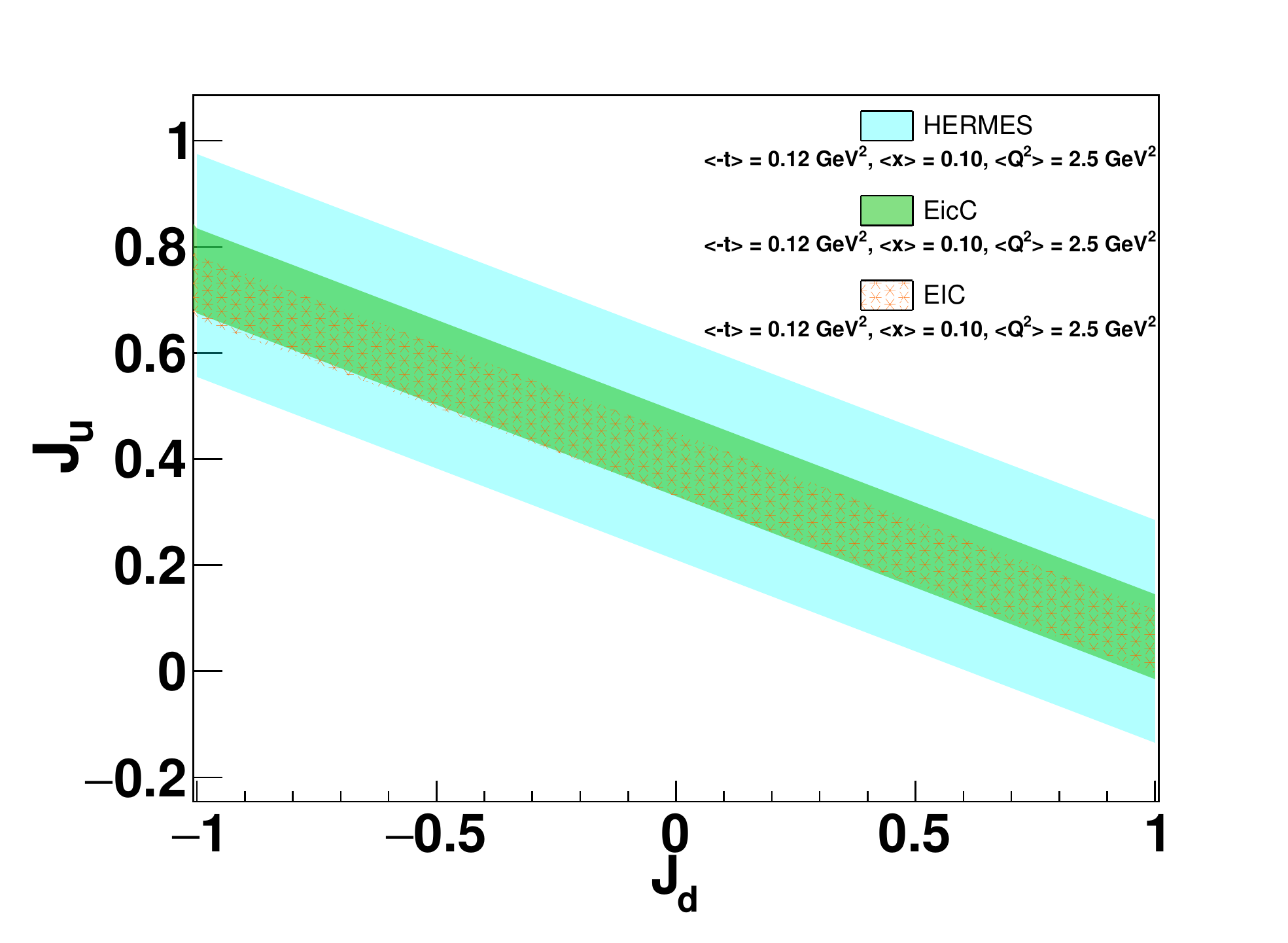}
	\caption{The result of model-dependent constraint on u-quark total angular momentum $J_u$ vs d-quark total angular momentum $J_d$ at EIC and EicC compared with HERMES \cite{Ellinghaus:2005uc,Ye:2006gza}. Both statistical and systematic errors are considered.}
	\label{fig:EIC_EicC_HERMESband01_sys}
\end{figure}

\begin{figure}[htbp]
	\centering
	\includegraphics[scale=0.46,angle=0]{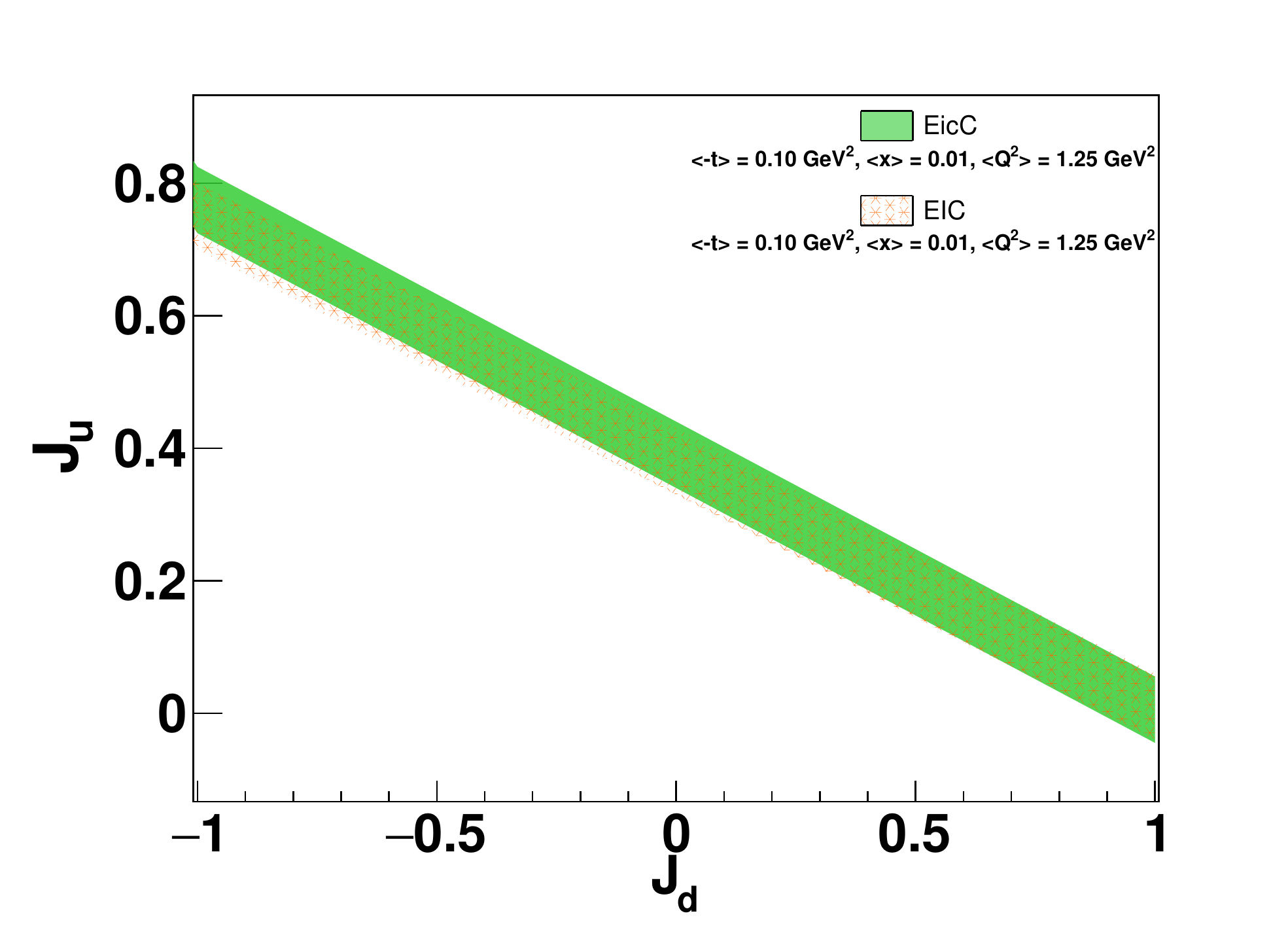}
	\caption{The result of model-dependent constraint on u-quark total angular momentum $J_u$ vs d-quark total angular momentum $J_d$ in the region of $x\sim0.01$ at EIC and EicC. Both statistical and systematic errors are considered.}
	\label{fig:EIC_EicC_band001_sys}
\end{figure}

\begin{figure}[htbp]
	\centering
	\includegraphics[scale=0.46,angle=0]{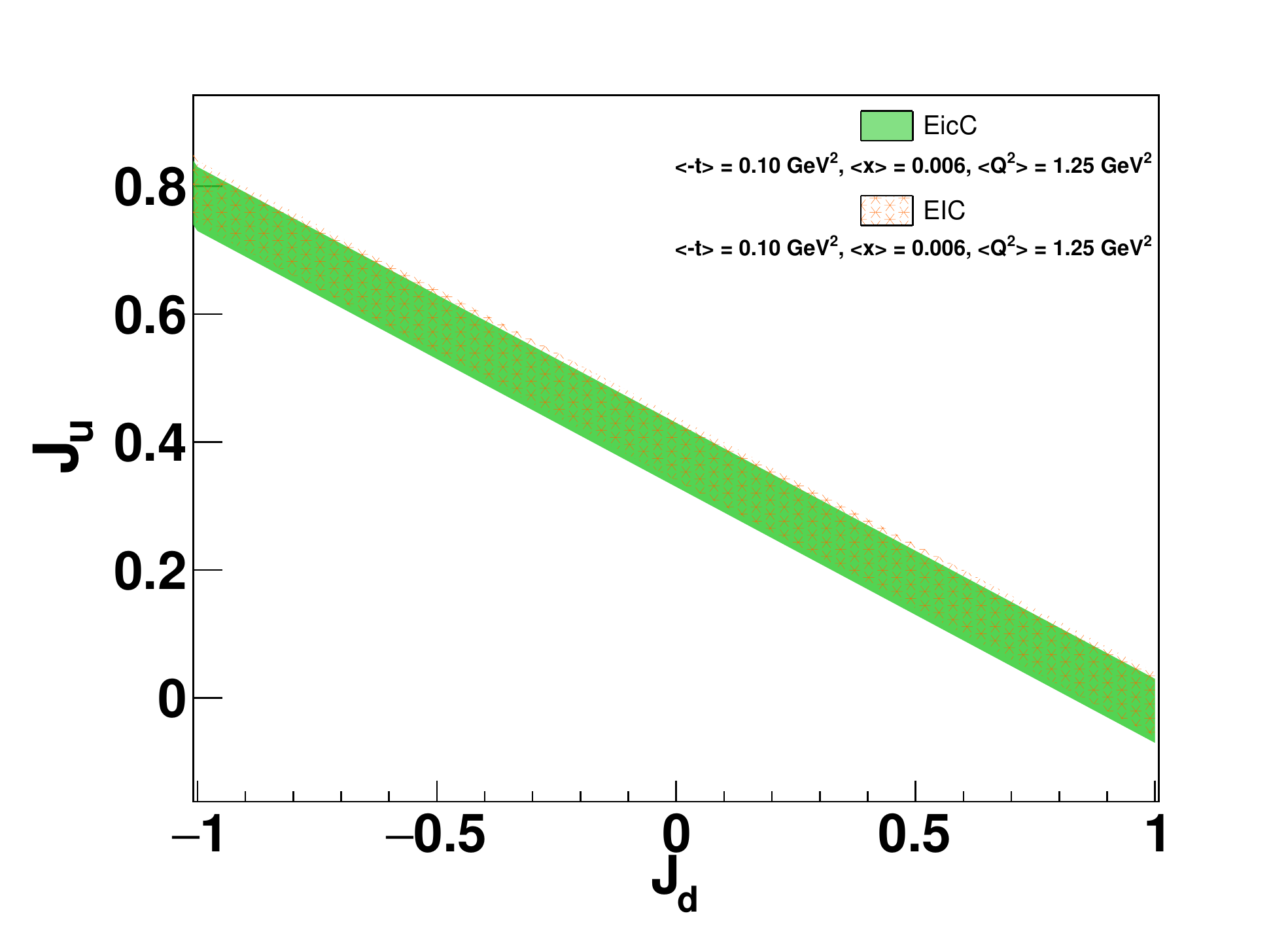}
	\caption{The result of model-dependent constraint on u-quark total angular momentum $J_u$ vs d-quark total angular momentum $J_d$ in the region of $x\sim0.006$ at EIC and EicC. Both statistical and systematic errors are considered.}
	\label{fig:EIC_EicC_band0006_sys}
\end{figure}

\begin{figure}[htbp]
	\centering
	\includegraphics[scale=0.46,angle=0]{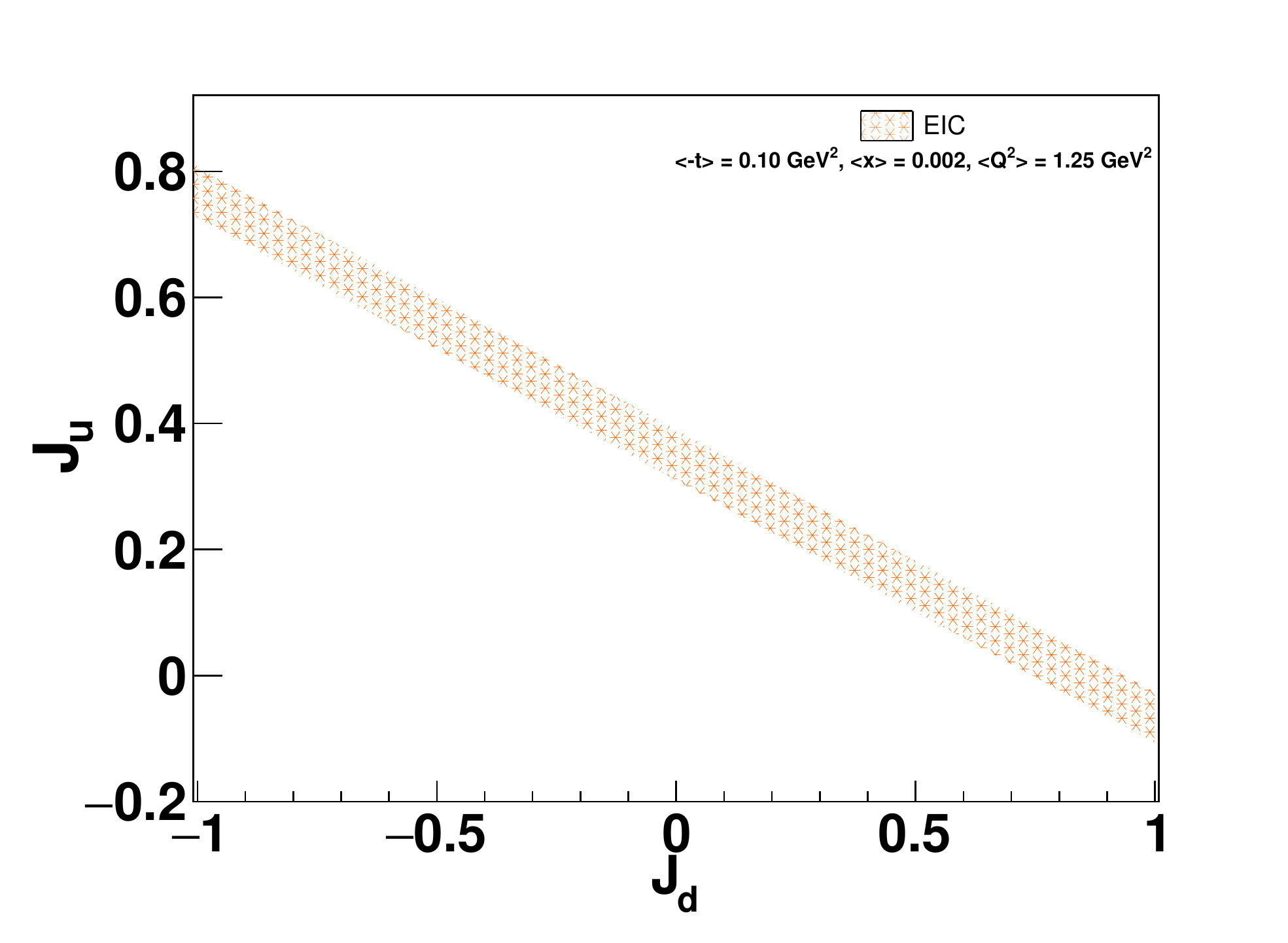}
	\caption{The result of model-dependent constraint on u-quark total angular momentum $J_u$ vs d-quark total angular momentum $J_d$ in the region of $x\sim0.002$ at EIC. Both statistical and systematic errors are considered.}
	\label{fig:EIC_band0002_sys}
\end{figure}

We develop a code to calculate observables in the exclusive reaction $e p \rightarrow e^{\prime} p^{\prime} \gamma$ to LO precision in perturbative theory. This calculation follows the VGG model described in Sec. \ref{sec:VGGmodel}. In order to compare the results from theoretical calculations with the TTSA amplitudes pseudo data in Fig. \ref{fig:Asystat_EicC_EIC_small_x}, the $\chi_{\text {exp }}^{2}$ is defined as:
\begin{equation}
	\begin{array}{l}
		\chi_{\exp }^{2}\left(J_{u}, J_{d}\right)=  \\
		\frac{\left[\left.A_{U T}^{\sin \left(\phi-\phi_{s}\right) \cos \phi}\right|_{(\text { Pseudo data })}-\left.A_{U T}^{\sin \left(\phi-\phi_{s}\right) \cos \phi}\right|_{\text {theory }}\right]^{2}}{\delta A_{\text {stat }}^{2} + \delta A_{\text {syst }}^{2}}.
	\end{array}
\end{equation}
There we need to consider the systematic errors. Based on the previous experiments \cite{H1:2001nez,ZEUS:2003pwh,H1:2005gdw,H1:2007vrx,H1:2009wnw,HERMES:2006pre,HERMES:2011bou,CLAS:2006krx,CLAS:2007clm,CLAS:2008ahu,JeffersonLabHallA:2006prd,JeffersonLabHallA:2007jdm,CLAS:2015bqi,Benali:2020vma,JeffersonLabHallA:2022pnx,JeffersonLabHallA:2015dwe,Defurne:2017paw,CLAS:2001wjj,CLAS:2015uuo,CLAS:2014qtk,CLAS:2018bgk,COMPASS:2018pup,Joerg:2016hhs}, we make a conservative estimate for EicC and EIC. Thus, for EicC and EIC, we assume experimental systematic errors are 10 $\%$.
The constraints on $J_u$ and $J_d$ obtained for the extracted TTSA amplitudes from the
pseudo data are shown in Fig. \ref{fig:Asystat_EicC_EIC_small_x}.
We calculate the TTSA amplitudes for $J_u$
($J_d$) ranging from 0 to 1 (-1 to 1) in steps of 0.2, and set the D-term = 0 ($D^{q}\left(\frac{x}{\xi}\right)$ in Eq. \ref{eq:profun}). 
Fig. \ref{fig:EIC_EicC_HERMESband01} shows the model-dependent constraint on u-quark total angular momentum $J_u$ vs d-quark total angular momentum $J_d$ in the same kinematic region as HERMES \cite{Ellinghaus:2005uc,Ye:2006gza}. 
Here we only consider the influences from statistical errors.
The result of EicC, which is shown in Fig. \ref{fig:EIC_EicC_HERMESband01}, can be expressed as
\begin{equation}
	J_u + J_d/2.9 = 0.41 \pm 0.06,
\end{equation}
and the result of EIC is
\begin{equation}
	J_u + J_d/3.0 = 0.39 \pm 0.04.
\end{equation}
If we consider both statistical and systematic errors ($A_{U T}^{\sin (\phi-\phi_s) \cos \phi} = -0.142 \pm 0.020 \pm 0.014$ at EicC, $A_{U T}^{\sin (\phi-\phi_s) \cos \phi} = -0.020 \pm 0.002 \pm 0.002$ at EIC), the result (shown in Fig. \ref{fig:EIC_EicC_HERMESband01_sys}) is
\begin{equation}
	J_u + J_d/2.9 = 0.41 \pm 0.08,
\end{equation}
for EicC, and
\begin{equation}
	J_u + J_d/3.0 = 0.39 \pm 0.06.
\end{equation}
for EIC. The uncertainty is propagated from the TTSA amplitudes uncertainty of the pseudo data, and experimental systematic errors dominate.
According to the results of HERMES \cite{Ellinghaus:2005uc,Ye:2006gza,HERMES:2008abz}, 
\begin{equation}
	J_u + J_d/2.9 = 0.42 \pm 0.21,
\end{equation}
we ignore the effects of parameter b and D-term. As the Fig. \ref{fig:EIC_EicC_HERMESband01_sys} shows, EicC and EIC have higher accuracy to obtain smaller uncertainty for constraint on u-quark and d-quark total angular momentum. 
Since EIC and EicC can provide a large amount of accurate data in the small $x$ region, we performed some calculations in this region. Both statistical and systematic errors are considered in these results.
At $x = 0.01$, the results of EicC and EIC are shown in Fig. \ref{fig:EIC_EicC_band001_sys}, where EicC is
\begin{equation}
	J_u + J_d/2.6 = 0.39 \pm 0.05,
\end{equation}
and EIC is
\begin{equation}
	J_u + J_d/2.7 = 0.38 \pm 0.05.
\end{equation}
In the smallest $x$ area that EicC can provide, we obtained the flowing results, where 
\begin{equation}
	J_u + J_d/2.5 = 0.38 \pm 0.05,
\end{equation}
is the result of EicC shown in Fig. \ref{fig:EIC_EicC_band0006_sys}. The result of EIC in this kinematic region is
\begin{equation}
	J_u + J_d/2.5 = 0.39 \pm 0.05.
\end{equation}
As Fig. \ref{fig:EicC_EIC_region} shows, EIC also provides accurate data in the area of $x\sim0.002$. In this very small $x$ region, we present the result of EIC, 
\begin{equation}
	J_u + J_d/2.4 = 0.35 \pm 0.04,
\end{equation}
which is shown in Fig. \ref{fig:EIC_band0002_sys}.

The results of EicC and EIC are both within the error range of HERMES and both have small errors. Without precise experiments, it is difficult for theoretical work to move forward. These precise experimental data will help us gain a deeper understanding of the nucleon structure in the future.

\section{Discussions and summary}
\label{sec:summary}

The internal structure of the nucleon is mysterious, and we explore it by various methods. After the EMC experiment, the researchers conducted many detailed studies of nucleon spins. 
The proposed GPDs theory opens new paths for the study of the three-dimensional structure and spin of nucleon. 
By Ji's sum rule, we find that GPDs are directly related to the total angular momentum carried by the partons. 
DVCS experiments are a good choice to obtain GPDs, although not quite directly extracted.
In contrast to the great progress in studying GPDs on the theoretical side, relatively little progress has been made on the experimental side. 
Because the experiment requires high statistical accuracy, this means that extremely good detectors and very high luminosity are required. 

In this work, we simulated the DVCS process at EicC and EIC to study the internal structure of proton. The statistical errors of these two future experiments are predicted.
According to the very small statistical errors, we find that the measurement
accuracy of future DVCS experiment will be limited mostly by systematical errors. 
It seems that the accuracy of the EIC and EicC data will be greatly improved in the future when compared with the existing real data from different experiment groups. 
Advanced experiment equipment to reduce systematic errors and better detection of final state particles to reduce statistical errors. We believe that future EicC and EIC experiments will yield more accurate data than those predicted in this work.
This has significant implications for future experimental studies of the internal structure of nucleon.
With the excellent detectors and high accelerator luminosity, DVCS experiments at EicC and EIC will have a bright prospect.

Based on the EIC and EicC measurements of TTSA high-precision pseudo-data, we can have a good study of the
nucleon helicity-flip GPD $E$. Through the VGG model, the GPD E is parameterized by the total angular momentum of the up and down quarks in the nucleon. With this model we combine DVCS experiments with nucleon spin studies.
According to the HERMES and JLab experiments constraint on the total angular momentum of quarks in the proton and neutron, 
we constraint on the total angular momentum carried by up quarks and down quarks inside the proton in future EIC and EicC experiments. 
There are different GPD models based on experimental and theoretical research to study the mysterious nucleon structure. Current research relies on models too heavily, we look forward to more precise experimental data to verify these theoretical research in the future.

\begin{acknowledgments}
	We thank Prof. J. P. Chen and Dr. Korotkov for suggestions and
	discussions. This work is
	supported by the Strategic Priority Research Program of
	Chinese Academy of Sciences under the Grant NO. XDB34030301.
\end{acknowledgments}

\bibliographystyle{apsrev4-1}
\bibliography{refs}

\end{document}